\documentclass[prb,twocolumn,showpacs,amsmath,amssymb]{revtex4}

\usepackage{graphicx}
\usepackage{dcolumn}
\usepackage{bm}
\usepackage{amsmath}

\begin{document}
\title
{Conserving Gapless Mean-Field Theory for Weakly Interacting Bose Gases}

\author{Takafumi Kita}
\affiliation{Division of Physics, Hokkaido University, Sapporo 060-0810, Japan}

\date{\today}

\begin{abstract}
This paper presents a conserving gapless mean-field theory 
for weakly interacting Bose gases.
We first construct a mean-field Luttinger-Ward thermodynamic functional
in terms of the condensate wave function $\Psi$ and
the Nambu Green's function $\hat{G}$ for the quasiparticle field.
Imposing its stationarity respect to $\Psi$ and $\hat{G}$
yields a set of equations to determine 
the equilibrium for general non-uniform systems.
They have a plausible property of satisfying
the Hugenholtz-Pines theorem to provide a gapless
excitation spectrum.
Also, the corresponding dynamical equations of motion 
obey various conservation laws.
Thus, the present mean-field theory shares two important properties
with the exact theory: ``conserving'' and ``gapless.''
The theory is then applied to a homogeneous
weakly interacting Bose gas 
with $s$-wave scattering length $a$ and particle mass $m$
to clarify its basic thermodynamic properties 
under two complementary conditions of constant density $n$
and constant pressure $p$.
The superfluid transition is predicted to be first-order because of 
the non-analytic nature of the order-parameter expansion near $T_{c}$ 
inherent in Bose systems, i.e., the
Landau-Ginzburg expansion is not possible here.
The transition temperature $T_{c}$ shows quite a different
interaction dependence between the $n$-fixed and $p$-fixed cases.
In the former case $T_{c}$ increases from the ideal gas value $T_{0}$
as $T_{c}/T_{0}\!=\! 1+ 2.33 an^{1/3}$,
whereas it decreases in the latter
as $T_{c}/T_{0}\!=\! 1\!-\! 3.84a(mp/2\pi\hbar^{2})^{1/5}$.
Temperature dependences of basic thermodynamic quantities are 
clarified explicitly.
\end{abstract}
\pacs{03.75.Hh, 05.30.Jp, 67.40.-w}

\maketitle

\section{Introduction}
\label{sec:intro}

Mean-field theories with self-consistency equations have
played a central role in the development of our microscopic understanding
on quantum many-particle systems, especially on broken-symmetry phases.
One of the most outstanding examples is undoubtedly
the Bardeen-Cooper-Schrieffer (BCS) theory of superconductivity,\cite{BCS57}
which may be regarded now as the 
Hartree-Fock theory in the Nambu space.\cite{Nambu60}
In some notable cases like the BCS theory,
mean-field theories
have brought remarkable quantitative descriptions
of experiments.
At least, each mean-field theory have provided a basic starting point
for later developments, as
the Weiss\cite{Weiss07} and the Stoner\cite{Stoner38} 
theories for localized and
itinerant ferromagnets, respectively.
Hence there is every need for a mean-field
theory of every quantum system.
However, there has been no established mean-field theory
for Bose-Einstein condensates (BEC).

Over the last sixty years, the field theory has manifested itself 
as one of the most powerful microscopic approaches to quantum 
many-particle systems.
It was first applied to Bose particles in the celebrated work
of Bogoliubov,\cite{Bogoliubov47}
followed by intensive theoretical investigations in later 
years.\cite{HY57,HYL57,BS57,BS57-2,LHY57,LY58,Beliaev58,ZT58,GA59,LY59,HP59,
Wentzel60,Takano61,Gross61,Pitaevskii61,Luban62,Shohno64,
GN64,DM64,HM65,PF65,Popov65,LG66,RS69,Fetter72}
However, it has brought a rather poor description of 
condensed Bose particles at finite temperatures.
For example, we still do not have a complete agreement
on the Bose-Einstein condensation temperature $T_c$ of 
the homogeneous weakly interacting Bose 
gas under constant density
$n$,\cite{BS96,GCL97,Holzmann99,Baym00,Arnold00,Arnold01,
Kashurnikov01,DM03,Kleinert03,Kastening04,LHK04,Andersen04,Yukalov04}
not to mention its thermodynamic properties 
over $0\!\leq\! T\!\leq\! T_{c}$;
see Refs.\ \onlinecite{Andersen04} and \onlinecite{Yukalov04}
for a review on the $T_c$ calculations.
This situation is partly due to the absence of 
a well-established renormalized
perturbation theory for Bose systems corresponding to
the Luttinger-Ward theory on Fermi liquids.\cite{LW60}
Indeed, important formal results on Bose systems have often
been obtained with the
bare perturbation theory, as in the case of 
the Hugenholtz-Pines theorem.\cite{HP59}
However, the bare perturbation expansion itself is not very suitable
for practical calculations due to the infrared divergences
inherent in Bose systems. 
On the other hand, we still do not have a systematic approximation scheme
of how to renormalize the condensate wave function and 
the quasiparticle Green's function self-consistently without losing
the physical essentials of condensed Bose particles.

The self-consistent Wick decomposition procedure has proved to be
quite powerful for most quantum many-particle systems
at finite temperatures
and might be used also for Bose particles.
The corresponding mean-field theory for condensed Bose systems
is known now as the Girardeau-Arnowitt theory\cite{GA59,HM65} or the
Hartree-Fock-Bogoliubov (HFB) theory.\cite{Griffin96}
However, it predicts an energy gap in the excitation spectrum,
in contradiction to the
Hugenholtz-Pines theorem of declaring a gapless excitation.
Thus, it has become customary to introduce a further approximation
to the HFB theory,
now generally called the ``Popov'' approximation,\cite{Griffin96}
of completely ignoring the anomalous quasiparticle pair correlation
to recover a gapless excitation.
When it is adopted to describe dynamics, however, the same approximation
does not satisfy various conservation laws as required.
Also, it is not clear whether it is permissible or not
to neglect the pair correlation completely.
Recently, Proukakis {\em et al}.\ \cite{Proukakis98}
have presented an improved gapless theory
with finite pair correlation. However,
it still does not satisfy the conservation laws.
Thus, there has been no mean-field theory 
for condensed Bose systems that simultaneously carries the
important properties of the exact theory pointed out by 
Hohenberg and Martin:\cite{HM65} 
``conserving'' and ``gapless.''

On the basis of these observations, we have recently formulated
a new mean-field theory for BEC with the desired 
conserving gapless character.\cite{Kita05}
We here present a detailed description of the theory with several new results.
As first shown by Baym for normal Fermi liquids\cite{Baym62} and
pointed out by Hohenberg and Martin for condensed Bose systems,\cite{HM65}
various conservation laws are automatically satisfied in 
the ``$\Phi$ derivative approximations''
where the irreducible self-energy $\hat{\Sigma}$ is obtained 
as a derivative of some functional
$\Phi\!=\!\Phi[\hat{G}]$ with respect to the renormalized 
Green's function $\hat{G}$.
This relationship between $\hat{\Sigma}$ and
$\hat{G}$ holds exactly 
for Fermi systems in equilibrium, as shown by Luttinger and Ward.\cite{LW60}
Indeed, $\Phi$ was first introduced by Luttinger and Ward as part of the
exact thermodynamic functional $\Omega\!=\!\Omega[\hat{G}]$. 
It was later taken up by Baym\cite{Baym62} to present a criterion for
obtaining approximate dynamical equations with conservation laws.
It thus follows that dynamical equations obtained from
a Luttinger-Ward functional naturally obey the conservation laws.
We hence ask the heuristic question here: 
Can we construct a mean-field Luttinger-Ward thermodynamic functional for BEC
that also satisfies the Hugenholtz-Pines theorem?
This is indeed possible as will be shown below.
The predictions of the resulting mean-field theory on the homogeneous
weakly interacting Bose gas
will be presented later with an expression for $\Delta T_c$;
this application hopefully will provide a deeper insight
into the superfluidity and phase transition of still 
mysterious $^{4}$He.\cite{Keller69,Lipa96}
It should be noted that
the same idea has been adopted more recently
by Ivanov, Riek, and Knoll\cite{IRK05} (IRK) 
in a different context of the $O(N)$ model
to present an apparently different thermodynamic functional.
It will be shown, however, that their functional yields exactly the
same equilibrium thermodynamic properties on weakly interacting Bose gases
as the present one.

We finally provide three comments on terminology
to remove possible confusions in advance. First,
Gardiner\cite{Gardiner97} and Castin and Dum\cite{CD98}
have presented an alternative description of BEC,
which was called by Gardiner\cite{Gardiner97} 
as ``particle-number conserving Bogoliubov method.''
However, his terminology is completely different from the
present ``conserving'' and we call their formulations as
``number-fixed'' descriptions of BEC.
Indeed, his terminology is
connected more closely with the equilibrium thermodynamics 
to denote the fact that the particle number
${N}$ is chosen as an independent variable instead of 
the chemical potential $\mu$.
The two descriptions 
of using ${N}$ and $\mu$ are equivalent in the thermodynamic limit
except the fluctuation in the particle number.
It is worth pointing out that a
``number-fixed'' theory does not necessarily
provide ``conserving'' dynamical equations
in the present sense.
Second, the term ``gapless''
is relevant here to the single-particle excitation.
Indeed, the collective excitation has another story and 
is gapless even in the Girardeau-Arnowitt theory, 
as already shown by Takano.\cite{Takano61}
Third, it has been pointed out by Yukalov\cite{Yukalov04}
that the word ``Popov''
may not be suitable for the approximation of 
setting the anomalous pair amplitude equal to zero
in the HFB theory.
Indeed, this approximation was introduced by Shohno\cite{Shohno64} already
in 1964 and used later by Reatto and Straley.\cite{RS69}
We can also find the same approximation later in the work of
Baym and Grinstein on the $\sigma$ model.\cite{BG77}
We will call it as the ``Shohno'' approximation
instead of ``Popov'' following Reatto and Straley.\cite{RS69}

This paper is organized as follows.
Section II presents a mean-field free-energy functional 
for general non-uniform systems
and derives a closed set of equations to determine the thermodynamic
equilibrium.
Section III applies the formulation to the homogeneous 
weakly interacting Bose gas under constant density
to clarify its basic thermodynamic properties.
Section IV treats the same system under the complementary
condition of constant pressure.
Section V concludes the paper.
Appendix A summarizes Feynman rules for the perturbation expansion
with respect to the Nambu Green's function.
Appendix B provides a proof 
on the condensed Bose systems that the conservation laws are
obeyed in the $\Phi$ derivative approximations.
In Appendix C, the connection between the present and 
the IRK theories is clarified. 
Appendix D discusses the origin of a non-Hermitian eigenvalue
problem in condensed Bose systems.
Appendix E derives compact expressions for the equilibrium
thermodynamic functional and the entropy.
We put $k_{\rm B}\!=\! 1$ throughout.

\section{Mean-Field equations}

\subsection{Free-energy functional}

We express the field operator $\psi({\bf r})$
as a sum of the condensate wave function $\Psi({\bf r})$
and the quasiparticle field $\phi({\bf r})$ as
\begin{equation}
\psi({\bf r})=\Psi({\bf r})+\phi({\bf r}) \, .
\end{equation}
It is convenient to introduce the spinors:
\begin{subequations}
\label{spinors}
\begin{eqnarray}
\vec{\phi}\equiv\left[
\begin{array}{c}
\vspace{1mm}
\phi \\ \phi^{\dagger}
\end{array}
\right]  ,
\hspace{5mm}
\vec{\phi}^{\,\dagger}\equiv
[\, \phi^{\dagger} \,\, \phi\,] \, ,
\label{phi}
\end{eqnarray}
\begin{eqnarray}
\vec{\Psi}\equiv\left[
\begin{array}{c}
\vspace{1mm}
\Psi \\ \Psi^{*}
\end{array}
\right]  ,
\hspace{5mm}
\vec{\Psi}^{\dagger}\equiv
[\, \Psi^{*} \,\, \Psi\,] \, .
\label{Psi}
\end{eqnarray}
\end{subequations}
Using Eq.\ (\ref{phi}), we define our Matsubara 
Green's function in Nambu space\cite{Nambu60} as
\begin{eqnarray}
&&\hspace{-5mm}
\hat{G}({\bf r},{\bf r}';\tau)\equiv
-
\hat{\tau}_{3} \,\langle T_{\tau}
\vec{\phi}({\bf r},\tau)
\vec{\phi}^{\,\dagger}({\bf r}')\rangle
\nonumber \\
&&\hspace{11mm}
= T\sum_{n=-\infty}^{\infty} 
\hat{G}({\bf r},{\bf r}';z_{n})\,{\rm e}^{-z_{n}\tau}
\, ,
\label{hatG}
\end{eqnarray}
with $\hat{\tau}_{3}$ the third Pauli matrix
and $z_{n}\!\equiv\! 2\pi i n T $.
The factor $\hat{\tau}_{3}$
is usually absent in the definition of
Green's function.\cite{HM65,Griffin96} 
Introducing the factor brings an advantage that 
the eigenvalue problem
for $\hat{G}^{-1}$ is equivalent to the 
Bogoliubov-de Gennes (BdG) equation for the quasiparticles,
as seen below.
Thus, $\hat{G}$ can be put into a diagonal form
with respect to the eigenstates of the BdG equation.

The bare perturbation expansion for condensed Bose systems 
may also be performed quite compactly
in terms of the Nambu-Matsubara Green's function
of Eq.\ (\ref{hatG}).
The corresponding Feynman rules can be found easily from
those of the superconducting Fermi systems\cite{Kita96} 
with slight modifications, which are summarized in Appendix A.
Retaining the lowest-order diagrams in terms of the interaction
and renormalizing 
$\hat{G}_{0}\!\rightarrow\!\hat{G}$, 
one may express the free-energy functional of the
HFB theory in a Luttinger-Ward form\cite{LW60,Kita96}
with respect to $\Psi$ and $\hat{G}$.

We now present our mean-field Luttinger-Ward functional
$\Omega\!=\!\Omega (\Psi,\Psi^{*},\hat{G})$ which
has been obtained from that of the HFB theory
with a slight modification.
It is given by
\begin{eqnarray}
&&\hspace{-7mm}
\Omega = 
\int \! \Psi^{*}({\bf r})K \Psi({\bf r}) 
\, d{\bf r}+
\frac{T}{2} \sum_{n} {\rm Tr}\!\left[
\ln \bigl(\hat{\tau}_{3}K\!+\!\hat{\Sigma} 
\!-\! z_{n}\hat{1}\bigr)\right.
\nonumber \\
&& \hspace{0mm}
\left.+\,\hat{G}
\hat{\Sigma}\,\right]\!\hat{1}(z_{n})
+ \Phi
\, .
\label{LWF}
\end{eqnarray}
Here $K$ is defined by
\begin{equation}
K\equiv 
-\frac{\hbar^{2}}{2m}\mbox{\boldmath $\nabla$}^{2}+{\cal V}({\bf r})-\mu_{\rm s} \, ,
\label{H^(0)}
\end{equation}
with ${\cal V}({\bf r})$ the external potential and 
$\mu_{\rm s}$ the chemical potential,
$\hat{1}$ is the unit matrix, and $\hat{1}(z_{n})$ denotes
\begin{equation}
\hat{1}(z_{n})
=\left[
\begin{array}{cc}
\vspace{1mm}
{\rm e}^{z_{n}0_{+}} & 0
\\
0 & {\rm e}^{-z_{n}0_{+}}
\end{array}
\right] ,
\label{1_n}
\end{equation}
with $0_{+}$ an infinitesimal positive constant.
The branch cut of the logarithm in Eq.\ (\ref{LWF}) is 
chosen along the negative real axis, and
Tr here includes an integration 
over space variables
with multiplications of $\hat{\tau}_{3}K$ and $-z_{n}\hat{1}$
by $\delta({\bf r}\!-\!{\bf r}')$ from the right implied.
Finally, $\hat{\Sigma}$ denotes the irreducible self-energy
obtained from the functional 
$\Phi\!=\!\Phi(\Psi,\Psi^{*},\hat{G})$ by
\begin{equation}
\hat{\Sigma}({\bf r},{\bf r}';z_{n})
\!=\!-\frac{2}{T}\,\frac{\delta \Phi}
{\delta \hat{G}({\bf r}',{\bf r};z_{n})} \, .
\label{SigmaDef}
\end{equation}
With Eq.\ (\ref{SigmaDef}), $\Omega$ becomes stationary with respect to
a variation in $\hat{G}$ satisfying Dyson's equation:
\begin{equation}
\hat{G}^{-1}= z_{n}\hat{1}-\hat{\tau}_{3}K-\hat{\Sigma} \, .
\label{Dyson}
\end{equation}
The condensate wave function $\Psi({\bf r})$
in equilibrium is also determined by
$\delta \Omega/\delta \Psi^{*}({\bf r})\!=\! 0$.
Noting $\delta \Omega/\delta \hat{G}\!=\! \hat{0}$,
we only need to consider the explicit $\Psi^{*}$ dependences
in $\Omega$ to obtain
\begin{equation}
K\Psi({\bf r}) = -\eta ({\bf r}) \, , 
\label{Psi-eta}
\end{equation}
with 
\begin{equation}
\eta ({\bf r})\equiv \frac{\delta \Phi}{\delta \Psi^{*}({\bf r})} \, .
\label{eta}
\end{equation}
The quantity $\eta ({\bf r})$ is the so-called
condensate source function.\cite{HM65,Griffin96}
Equations (\ref{SigmaDef}) and (\ref{eta}) constitute the $\Phi$
derivative approximation\cite{HM65} where
the conservation laws are obeyed automatically
by the dynamical equations.
This is one of the main advantages in using the Luttinger-Ward functional
as a starting point.

Despite their importance,
the conservation laws in the $\Phi$-derivative 
approximation seems not to have been described 
to an enough extent in the literature for the condensed Bose systems. 
We hence provide a detailed proof 
of them in Appendix B.

A key quantity in Eq.\ (\ref{LWF}) is $\Phi$.
We choose it so that
the Hugenholtz-Pines theorem is satisfied simultaneously.
Explicitly, it is given by
\begin{subequations}
\label{Phi-total}
\begin{eqnarray}
&&\hspace{-8mm}
\Phi =\!\int\! d{\bf r}\! \int\! d{\bf r}'\,
{\cal U}({\bf r}\!-\!{\bf r}')
\biggl[ \frac{1}{2}\, |\Psi({\bf r})|^{2}|\Psi({\bf r}')|^{2}
\nonumber \\
&& \hspace{-5mm}
-\frac{T}{2} \! \sum_{n}
|\Psi({\bf r})|^{2}\,
{\rm Tr}\, \hat{\tau}_{3}\hat{G}({\bf r}',{\bf r}';z_{n})
\hat{1}(z_{n})
\nonumber \\
&& \hspace{-5mm} 
-\frac{T}{2} \! \sum_{n}
{\rm Tr}\, \hat{\tau}_{3} 
\vec{\Psi}({\bf r})
\vec{\Psi}^{\dagger} ({\bf r}')
\hat{G}({\bf r}',{\bf r};z_{n})
\hat{1}(z_{n})
\nonumber \\ 
&& \hspace{-5mm}
+ \frac{T^{2}}{8} \! \sum_{n,n'} \!
{\rm Tr}\,\hat{\tau}_{3}\hat{G}({\bf r},{\bf r}, z_{n})
\hat{1}(z_{n})
{\rm Tr}\, \hat{\tau}_{3}\hat{G}({\bf r}',{\bf r}';z_{n'})
\hat{1}(z_{n'})
\nonumber \\
&& \hspace{-5mm}
+ \frac{T^{2}}{4} \! 
\sum_{n,n'}\! {\rm Tr}\, \hat{G}({\bf r},{\bf r}';z_{n})
\hat{1}(z_{n})
\hat{G}({\bf r}',{\bf r};z_{n'})
\hat{1}(z_{n'})\biggr] ,
\label{Phi}
\end{eqnarray}
where ${\cal U}$ denotes the interaction potential.
As may be realized from Appendix A
where the Feynman rules in Nambu space are given explicitly, 
the five terms in the square brackets of Eq.\ (\ref{Phi}) 
corresponds to the diagrams
of Fig.\ 1(a)-(e), respectively.
The difference of Eq.\ (\ref{Phi}) from $\Phi_{\rm HFB}$
lies in the Fock terms, i.e., the third and the fifth terms. 
Indeed, $\Phi_{\rm HFB}$\cite{HM65,Griffin96} is recovered
from Eq.\ (\ref{Phi})
by replacing $\hat{G}$ and
$\hat{\tau}_{3} 
\vec{\Psi}({\bf r})
\vec{\Psi}^{\dagger} ({\bf r}')$ 
in the two Fock terms by
$\hat{\tau}_{3} \hat{G}$ and
$\vec{\Psi}({\bf r})
\vec{\Psi}^{\dagger} ({\bf r}')$, respectively;
the present functional was found in the reverse way.

\begin{figure}[t]
\begin{center}
\includegraphics[width=7cm]{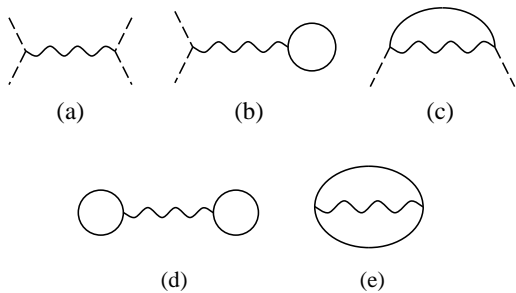}%
\end{center}
\caption{Diagrams contributing to $\Phi$.
The wavy, the solid, and the broken lines denote
the interaction, the Nambu Green's function,
and the condensate wave function, respectively.}
\label{fig:1}
\end{figure}

At this stage, it may be worth providing a comment 
on the present functional.
As mentioned by Shi and Griffin,\cite{SG98}
the HFB approximation yields over-counting 
of the off-diagonal self-energy
diagrams in the bare perturbation expansion, 
thereby leading to an unphysical energy gap in the excitation spectrum.
Hence it is necessary to subtract the extra
contributions from the HFB self-energy.
In this context, it is interesting to note that 
the present functional removes some of the off-diagonal
terms in $\Phi_{\rm HFB}$ as
\begin{eqnarray}
&&\hspace{-9mm}
\Phi = \Phi_{\rm HFB}-\!\int\!\! d{\bf r}\!\!
\int\!\!d{\bf r}'
{\cal U}({\bf r}\!-\!{\bf r}')\bigl[\Psi({\bf r})\Psi({\bf r}')
\langle\phi^{\dagger}({\bf r}')
\phi^{\dagger}({\bf r})\rangle
\nonumber \\
&&\hspace{-3mm}
+\Psi^{*}({\bf r})\Psi^{*}({\bf r}')
\langle\phi({\bf r}')
\phi({\bf r})\rangle
+|\langle\phi({\bf r})
\phi({\bf r}')\rangle|^{2}\,\bigr] \, .
\label{Phi-PhiHFB}
\end{eqnarray}
\end{subequations}
It satisfies the Hugenholtz-Pines relation appropriately
to be free from the unphysical excitation gap,
as seen below. Thus, the subtraction mentioned above may have been
performed appropriately.

Based on exactly the same idea as adopted here,
Ivanov, Riek, and Knoll\cite{IRK05}
have recently constructed an alternative functional $\Phi_{\rm IRK}$ 
for the $O(N)$ model so as to satisfy the Hugenholtz-Pines theorem.
Their functional is given in the present context by 
\begin{equation}
\Phi_{\rm IRK} = \Phi_{\rm HFB}-\!\int\!\! d{\bf r}\!\!
\int\!\!d{\bf r}'
{\cal U}({\bf r}\!-\!{\bf r}')|\langle\phi({\bf r})
\phi({\bf r}')\rangle|^{2} \, .
\label{Phi-IRK}
\end{equation}
In spite of the apparent difference between Eqs.\ (\ref{Phi-PhiHFB}) and
(\ref{Phi-IRK}), however, 
$\Phi$ and $\Phi_{\rm IRK}$ lead to exactly the 
same thermodynamic properties
for weakly interacting Bose gases, 
as shown in Appendix C.

\subsection{Equilibrium solution}

Now that $\Phi$ is given explicitly, we obtain
the equilibrium self-energy by Eq.\ (\ref{SigmaDef}). 
It may be written as
\begin{equation}
\hat{\Sigma}({\bf r},{\bf r}')
=
\left[
\begin{array}{cc}
\vspace{1mm}
\Sigma({\bf r},{\bf r}') &\Delta({\bf r},{\bf r}')
\\
-\Delta^{\!*}({\bf r},{\bf r}') &-\Sigma^{*}({\bf r},{\bf r}')
\end{array}
\right] ,
\label{hatSigma}
\end{equation}
where $\Sigma$ and $\Delta$ are given by
\begin{subequations}
\label{SD}
\begin{eqnarray}
&& \hspace{-13mm} \Sigma({\bf r},{\bf r}')
 =\delta({\bf r}\!-\!{\bf r}') \!
\int \! {\cal U}({\bf r}\!-\!{\bf r}'')\rho({\bf r}'',{\bf r}'') \, 
d{\bf r}''
\nonumber \\
&&\hspace{3mm}
+ \, {\cal U}({\bf r}\!-\!{\bf r}')\rho({\bf r},{\bf r}') \, ,
\label{Sigma}
\end{eqnarray}
\begin{equation}
\Delta({\bf r},{\bf r}')
 = {\cal U}({\bf r}\!-\!{\bf r}')\tilde{\rho}({\bf r},{\bf r}') \, ,
\label{Delta}
\end{equation}
\end{subequations}
respectively, with
\begin{subequations}
\label{rho}
\begin{eqnarray}
&& \hspace{-13mm} \rho({\bf r},{\bf r}')
 =\Psi({\bf r})\Psi^*({\bf r}')+
\langle \phi^\dagger({\bf r}')\phi({\bf r})\rangle \, ,
\label{rho1}
\end{eqnarray}
\begin{equation}
\tilde{\rho}({\bf r},{\bf r}')
 = \Psi({\bf r})\Psi({\bf r}')+
\langle \phi({\bf r})\phi({\bf r}')\rangle \, .
\label{rho2}
\end{equation}
\end{subequations}
The equation for the condensate wave function in equilibrium is obtained
from Eqs.\ (\ref{Psi-eta}), (\ref{eta}), and 
(\ref{Phi}). It can be written explicitly 
in terms of Eq.\ (\ref{SD}) by
\begin{equation}
K\Psi({\bf r})+\int\!\left[
\Sigma({\bf r},{\bf r}')\Psi({\bf r}')\!-\!\Delta({\bf r},{\bf r}')
\Psi^{*}({\bf r}')\right] d{\bf r}'= 0\, .
\label{GP}
\end{equation}
In the homogeneous case with no external potential where
$K\!\rightarrow\!-\mu_{\rm s}$ and
$\Psi\!=\! \sqrt{n_{0}}$ with $n_{0}$ the condensate density,
Eq.\ (\ref{GP}) reduces to the
Hugenholtz-Pines relation 
$\mu_{\rm s}\!=\! \Sigma_{{\bf k}={\bf 0}}\!-\!\Delta_{{\bf k}={\bf 0}}$,
as desired.
The expression for the particle number 
${N}\!=\!-\partial\Omega/\partial \mu_{\rm s}$ 
is found similarly as 
\begin{equation}
{N}=\int \! \rho({\bf r},{\bf r})\, d{\bf r} ,
\label{N}
\end{equation}
with $\rho$ given by Eq.\ (\ref{rho1}).
Using this equation, one may change 
independent variables of the free energy as
$\Omega(T,V,\mu)\!\rightarrow\! F(T,V,{N})\!\equiv\!
\Omega(T,V,\mu)\!+\!\mu {N}$.
Equations (\ref{Dyson}) and (\ref{hatSigma})-(\ref{N})
determines the equilibrium.

\subsection{Bogoliubov-de Gennes equations}

Physical quantities in equilibrium may be calculated most conveniently
in the representation where $\hat{G}$ is diagonal.
Noting Eq.\ (\ref{Dyson}), we introduce
\begin{equation}
\hat{H}({\bf r},{\bf r}')\!\equiv\!
\hat{\tau}_{3}K\delta({\bf r}\!-\!{\bf r}')
\!+\!\hat{\Sigma}({\bf r},{\bf r}') \, .
\label{hatH}
\end{equation}
The eigenvalue problem for $\hat{H}$
constitutes the BdG equation:
\begin{subequations}
\begin{equation}
\int
\hat{H}({\bf r},{\bf r}')\vec{\varphi}_{\mu}({\bf r}')
\, d{\bf r}'
=E_{\mu} \vec{\varphi}_{\mu}({\bf r}) \, ,
\label{BdG-vec1}
\end{equation}
where $\vec{\varphi}_{\mu}$ denotes an eigenfunction.
It has a peculiar feature that $\hat{H}$ is not Hermitian
but satisfies $\hat{H}^{\dagger}({\bf r},{\bf r}')
\!=\!\hat{\tau}_{3}\hat{H}({\bf r}',{\bf r})\hat{\tau}_{3}$.
Since most of the eigenvalue problems in quantum mechanics are
Hermitian by nature, it may be worth tracing 
why we have to treat a non-Hermitian matrix in BEC.
It is shown in Appendix D that this symmetry necessarily results
from the requirement that the quasiparticle field obey the Bose
statistics.

Let us enumerate basic properties of the BdG equation,
since this seems not to have been performed to an enough extent
in the literature.
We will also provide a compact Nambu representation to 
every quantity of the BdG equation.
Taking its Hermitian conjugate and
using $\hat{H}^{\dagger}({\bf r},{\bf r}')
\!=\!\hat{\tau}_{3}\hat{H}({\bf r}',{\bf r})\hat{\tau}_{3}$,
Eq.\ (\ref{BdG-vec1}) may be written alternatively as
\begin{equation}
\int
\vec{\varphi}_{\mu'}^{\,\dagger}({\bf r})
\hat{\tau}_{3}\hat{H}({\bf r},{\bf r}')\hat{\tau}_{3}
d{\bf r}
=E_{\mu'}^{*} \vec{\varphi}_{\mu'}^{\,\dagger}({\bf r}') \, .
\label{BdG-vec2}
\end{equation}
\end{subequations}
Let us multiply Eq.\ (\ref{BdG-vec1}) by
$\vec{\varphi}_{\mu'}^{\,\dagger}({\bf r})
\hat{\tau}_{3}$ from the left and integrate it over ${\bf r}$.
Similarly, we multiply
Eq.\ (\ref{BdG-vec2}) by
$\hat{\tau}_{3}\vec{\varphi}_{\mu}({\bf r}')
$ from the right and integrate it over ${\bf r}'$.
Subtracting the latter from the former, we obtain
\begin{equation}
(E_{\mu}-E_{\mu'}^{*}) \langle
\vec{\varphi}_{\mu'}|\hat{\tau}_{3}|
\vec{\varphi}_{\mu}\rangle =0 \, .
\label{BdG-prop}
\end{equation}
We first put $\mu'\!=\!\mu$ in Eq.\ (\ref{BdG-prop}). 
We then realize that $E_{\mu}$ is real 
as long as 
$\langle\vec{\varphi}_{\mu}|\hat{\tau}_{3}|
\vec{\varphi}_{\mu}\rangle\!\neq\! 0$ 
can be satisfied by the eigenstate.
We assume that (i) this is the case and (ii) eigenstates
with positive eigenvalues may be normalized as 
$\langle\vec{\varphi}_{\mu}|\hat{\tau}_{3}|
\vec{\varphi}_{\mu}\rangle\!=\! 1$.
We next consider the case $E_{\mu'}\!\neq\!E_{\mu}$
in Eq.\ (\ref{BdG-prop}).
We then find that 
the two eigenstates with different eigenvalues
are orthogonal as $\langle\vec{\varphi}_{\mu'}|\hat{\tau}_{3}|
\vec{\varphi}_{\mu}\rangle\!=\! 0$.

The Hamiltonian $\hat{H}$ has another symmetry:
$\hat{H}({\bf r},{\bf r}')
\!=\!-\hat{\tau}_{1}\hat{H}^{*}({\bf r},{\bf r}')\hat{\tau}_{1}$
with $\hat{\tau}_{1}$ the first Pauli matrix.
Let us take the complex conjugate of Eq.\ (\ref{BdG-vec1}),
multiply it by $-\hat{\tau}_{1}$ from the left, and
use the above symmetry.
It is thereby shown that a positive eigenvalue $E_{\nu}$ 
of the BdG equation with the eigenfunction:
\begin{subequations}
\begin{equation}
\vec{u}_{\nu}({\bf r})\equiv
\left[
\begin{array}{c}
u_{\nu}({\bf r}) \\ -v_{\nu}^{*}({\bf r})
\end{array}
\right] ,
\end{equation}
is always accompanied by the negative eigenvalue $-E_{\nu}$
with the eigenfunction:
\begin{equation}
\vec{v}_{\nu}({\bf r})\equiv
\left[
\begin{array}{c}
-v_{\nu}({\bf r}) \\ u_{\nu}^{*}({\bf r})
\end{array}
\right] .
\end{equation}
The sign in front of $v_{\nu}^{*}$ is introduced 
for convenience to make the coefficient
$v_{\bf k}$ of the homogeneous system
positive.
As mentioned above, 
we assume that $\vec{u}_{\nu}({\bf r})$
with $E_{\nu}\!>\! 0$
can be normalized as 
\begin{equation}
\langle u_{\nu}|u_{\nu}\rangle\!-\!
\langle v_{\nu}|v_{\nu}\rangle \!=\! 1 \, .
\label{uv-norm}
\end{equation}
\end{subequations}
Violation of this condition marks 
an instability of the assumed $\Psi({\bf r})$,
as is the case of a vortex-free $\Psi({\bf r})$
under an angular velocity 
$\Omega$ beyond the critical
value $\Omega_{c1}$.

It also follows from Eq.\ (\ref{GP})
that the BdG equation has eigenvalue $0$ whose eigenfunction 
is proportional to $\hat{\tau}_{3}|\vec{\Psi}\rangle$
with $|\vec{\Psi}\rangle$ given by Eq.\ (\ref{Psi}).
We adopt the following normalized eigenfunction for this state:
\begin{subequations}
\label{2phi_0}
\begin{equation}
\vec{\varphi}_{0}^{\,(1)}\!({\bf r}) \equiv 
\frac{1}{\langle\vec{\Psi}|\vec{\Psi}\rangle^{1/2}}
\hat{\tau}_{3}\vec{\Psi}({\bf r})
=
\left[
\begin{array}{c}
\vspace{1mm}
\varphi_{0}({\bf r})
\\
-\varphi_{0}^{*}({\bf r})
\end{array}
\right] .
\label{phi_0}
\end{equation}
It is orthogonal to $| \vec{u}_{\nu}\rangle$ 
as $ \langle \vec{u}_{\nu}|\hat{\tau}_{3}|
\vec{\varphi}_{0}^{(1)}\rangle\!=\!0$, i.e., 
$\langle u_{\nu}|\varphi_{0}\rangle\!=\!
\langle \varphi_{0}|v_{\nu}\rangle$.
Noting Eq.\ (\ref{uv-norm}), we may assume
\begin{equation}
\langle u_{\nu}|\varphi_{0}\rangle=
\langle v_{\nu}|\varphi_{0}\rangle=0 \, .
\label{u-phi-norm}
\end{equation}
There still remains another independent state:
\begin{equation}
\vec{\varphi}_{0}^{\,(2)}\!({\bf r})
\equiv
\hat{\tau}_{3}\vec{\varphi}_{0}^{\,(1)}\!({\bf r})
 =
\left[
\begin{array}{c}
\vspace{1mm}
\varphi_{0}({\bf r})
\\
\varphi_{0}^{*}({\bf r})
\end{array}
\right] ,
\label{Tphi_0}
\end{equation}
\end{subequations}
which does not belong to $\hat{H}$, however.
The functions $\vec{\varphi}_{0}^{\,(j)}\!({\bf r})$ ($j\!=\! 1,2$)
satisfy 
$\langle \vec{\varphi}_{0}^{\,(i)}|
\hat{\tau}_{3}|\vec{\varphi}_{0}^{\,(j)}\rangle\!=\! 1\!-\!
\delta_{ij}$.

The functions $|\vec{u}_{\nu}\rangle$,
$|\vec{v}_{\nu}\rangle$, $|\vec{\varphi}_{0}^{\,(1)}\rangle$,
and $|\vec{\varphi}_{0}^{\,(2)}\rangle$ are assumed to form a complete set.
As shown by expanding an arbitrary spinor 
with respect to the basis functions and calculating 
the corresponding expansion coefficients,
this statement is equivalent to
\begin{eqnarray}
&&\hspace{-5mm}
\sum_{\nu}\bigl(|\vec{u}_{\nu}\rangle\langle \vec{u}_{\nu}|\hat{\tau}_{3}
- |\vec{v}_{\nu}\rangle\langle \vec{v}_{\nu}|\hat{\tau}_{3}\bigr)
\nonumber \\
&&\hspace{-5mm}
+|\vec{\varphi}_{0}^{\,(1)}\rangle
\langle \vec{\varphi}_{0}^{\,(2)}|\hat{\tau}_{3}
+|\vec{\varphi}_{0}^{\,(2)}\rangle
\langle \vec{\varphi}_{0}^{\,(1)}|\hat{\tau}_{3}= 
\hat{1} \, .
\label{complete-vec}
\end{eqnarray}

In summary, the BdG equation
for a pair of eigenstates $\pm E_{\nu}$ 
($E_{\nu}\!>\! 0$) can be written 
compactly in a matrix form as
\begin{equation}
\int \hat{H}({\bf r},{\bf r}')\hat{u}_{\nu}({\bf r}')
d{\bf r}'
= \hat{u}_{\nu}({\bf r}) 
E_{\nu} \hat{\tau}_{3} \, ,
\label{BdG}
\end{equation}
where $\hat{u}_{\nu}$ is defined by
\begin{subequations}
\label{basisHat}
\begin{equation}
\hat{u}_{\nu}({\bf r})=
\left[
\begin{array}{cc}
\vspace{1mm}
u_{\nu}({\bf r}) & -v_{\nu}({\bf r})
\\
-v_{\nu}^{*}({\bf r}) & u_{\nu}^{*}({\bf r})
\end{array}
\right] \, .
\label{uHat}
\end{equation}
There exists another independent state
composed of the condensate wave function $\Psi({\bf r})$:
\begin{equation}
\hat{\varphi}_{0}({\bf r})=
\left[
\begin{array}{cc}
\vspace{1mm}
\varphi_{0}({\bf r}) 
& \varphi_{0}({\bf r})
\\
-\varphi_{0}^{*}({\bf r}) 
& \varphi_{0}^{*}({\bf r})
\end{array}
\right]  , \hspace{5mm} 
\varphi_{0}({\bf r})=\frac{1}{\sqrt{2N_{0}}}\Psi({\bf r}) \, ,
\label{HatVarphi}
\end{equation}
\end{subequations}
with $N_{0}\!\equiv\!\frac{1}{2}
\langle\vec{\Psi}|\vec{\Psi}\rangle$
denoting the condensate number.
The orthonormality of the basis functions reads
\begin{subequations}
\label{ortho}
\begin{equation}
\int \hat{\tau}_{3}\hat{u}_{\nu}^{\dagger}({\bf r})\hat{\tau}_{3}
\hat{u}_{\nu'}({\bf r}) d{\bf r}= \delta_{\nu\nu'}\hat{1} \, ,
\label{ortho1}
\end{equation}
\begin{equation}
\int \hat{\tau}_{1}
\hat{\varphi}_{0}^{\dagger}({\bf r})\hat{\tau}_{3}
\hat{\varphi}_{0}({\bf r}) d{\bf r}= \hat{1} \, ,
\label{ortho2}
\end{equation}
\begin{equation}
\int \hat{\tau}_{3}\hat{u}_{\nu}^{\dagger}({\bf r})\hat{\tau}_{3}
\hat{\varphi}_{0}({\bf r}) d{\bf r}= \hat{0} \, .
\label{ortho3}
\end{equation}
\end{subequations}
The completeness (\ref{complete-vec}) can be written alternatively as
\begin{equation}
\sum_{\nu}
\hat{u}_{\nu}({\bf r})\hat{\tau}_{3}\hat{u}_{\nu}^{\dagger}({\bf r}')
\hat{\tau}_{3}
+ \hat{\varphi}_{0}({\bf r})\hat{\tau}_{1}
\hat{\varphi}_{0}^{\dagger}({\bf r}')
\hat{\tau}_{3}=
\delta({\bf r}-{\bf r}')\hat{1} \, ,
\label{complete}
\end{equation}
as seen by writing down the matrix elements explicitly.

Equations (\ref{BdG})-(\ref{complete}) are the basic properties of
the eigenstates of the BdG equation.
Noting that $\vec{\varphi}_{0}^{\,(2)}\!({\bf r})$ is
absent in $\hat{H}$, Eq.\ (\ref{hatH}) may be expanded as
\begin{equation}
\hat{H}({\bf r},{\bf r}')=\sum_{\nu}
\hat{u}_{\nu}({\bf r})E_{\nu}\hat{\tau}_{3}\hat{\tau}_{3}
\hat{u}_{\nu}^{\dagger}({\bf r}')
\hat{\tau}_{3} \, .
\label{hatH-exp}
\end{equation}
Green's function 
$\hat{G}\!=\!(z_{n}\hat{1}\!-\!\hat{H})^{-1}$
becomes diagonal in the representation
where $\hat{H}$ is. In addition, eigenvalue $0$
is absent in $\hat{G}$. Hence $\hat{G}({\bf r},{\bf r}';z_{n})$
can be written as
\begin{subequations}
\label{hatG-2}
\begin{equation}
\hat{G}({\bf r},{\bf r}';z_{n})=\sum_{\nu}
\hat{u}_{\nu}({\bf r})\hat{G}_{\nu}(z_{n})
\hat{\tau}_{3}\hat{u}_{\nu}^{\dagger}({\bf r}')\hat{\tau}_{3} \, ,
\label{Gexp}
\end{equation}
with
\begin{equation}
\hat{G}_{\nu}(z_{n}) = \left[
\begin{array}{cc}
\vspace{2mm}
(z_{n}\!-\! E_{\nu})^{-1} & 
0
\\
0 & 
(z_{n}\!+\! E_{\nu})^{-1}
\end{array}
\right] .
\label{G_nu}
\end{equation}
\end{subequations}

Substituting Eq.\ (\ref{hatG-2}) into Eq.\ (\ref{hatG})
and performing summation over $z_{n}$ with $\tau\!=\! -0_{+}$,
we obtain the expressions for
$\langle\phi^{\dagger}({\bf r}')\phi({\bf r})\rangle$ 
and 
$\langle\phi({\bf r})\phi({\bf r}')\rangle$ as
\begin{subequations}
\label{GF}
\begin{equation}
\langle\phi^{\dagger}({\bf r}')\phi({\bf r})\rangle\!=\!
\sum_{\nu}\left[u_{\nu}({\bf r})u_{\nu}^{*}({\bf r}')n_{\nu}\!+\!
v_{\nu}({\bf r})v_{\nu}^{*}({\bf r}')(1\!+\!n_{\nu})\right]  ,
\label{G}
\end{equation}
\begin{equation}
\langle\phi({\bf r})\phi({\bf r}')\rangle\!=\!\frac{1}{2}
\sum_{\nu}\left[u_{\nu}({\bf r})v_{\nu}({\bf r}')\!+\!
v_{\nu}({\bf r})u_{\nu}({\bf r}')\right](1\!+\!2n_{\nu})
 ,
\label{F}
\end{equation}
\end{subequations}
where $n_{\nu}\!\equiv
\!({\rm e}^{E_{\nu}/T}\!-\! 1)^{-1}$
is the Bose distribution function,
and we have used the $(1,2)$ element of Eq.\ (\ref{complete})
to make Eq.\ (\ref{F}) manifestly symmetric
with respect to ${\bf r}$ and ${\bf r}'$.

Equations (\ref{SD})-(\ref{hatH}), (\ref{BdG}), and (\ref{GF}) 
form a closed 
set of self-consistent equations satisfying
both the Hugenholtz-Pines theorem
and various conservation laws.
Note that the pair correlation $\langle\phi\phi\rangle$
is adequately included in Eqs.\ (\ref{GP}) and (\ref{BdG});
neglecting this contribution
yields the Shohno theory.\cite{Shohno64}

\subsection{Expressions of equilibrium}

It is shown in Appendix E
that we can transform Eq.\ (\ref{LWF}) 
in equilibrium into
\begin{eqnarray}
&&\hspace{-8mm}
\Omega_{\rm eq} = T \sum_{\nu} 
\ln(1\!-\! {\rm e}^{-E_{\nu}/T})-  \sum_{\nu} 
E_{\nu}\int\! |v_{\nu}({\bf r})|^{2}\, d{\bf r}
\nonumber \\
&& \hspace{0mm}
- \frac{1}{2}\! \int \! d{\bf r}\!\int \! d{\bf r}'\biggl[
\Sigma({\bf r},{\bf r}')\rho({\bf r}',{\bf r})
\!-\! \Delta^{\!*}({\bf r},{\bf r}')\tilde{\rho}({\bf r}',{\bf r})\biggr]
\, .
\nonumber \\
\label{Omega-eq}
\end{eqnarray}
The entropy in equilibrium can be obtained from
$S_{\rm s}\!=\!-\partial \Omega/\partial T$
by differentiating Eq.\ (\ref{LWF})
in terms of the explicit 
$T$ dependences. As shown in Appendix E, 
this yields
\begin{equation}
S_{\rm s}=\sum_{\nu}\left[(1\!+\! n_{\nu})\ln (1\!+\! n_{\nu})
-n_{\nu}\ln n_{\nu}\right]\, .
\label{S}
\end{equation}
This completes the formulation of our mean-field theory.

\section{Homogeneous gas under constant density}

\subsection{Equations of equilibrium}

We now apply the previous formalism to 
a homogeneous weakly interacting Bose gas with 
volume ${V}$ and particle number ${N}$.
The interaction we adopt is given by
\begin{equation}
{\cal U}({\bf r}\!-\!{\bf r}')\!=\! 
\frac{4\pi\hbar^{2} a}{m}
\delta({\bf r}\!-\!{\bf r}') \, ,
\label{V}
\end{equation}
where $m$ is the particle mass
and $a$ is the $s$-wave scattering length.
The important dimensionless parameter of the system is given by
\begin{equation}
\delta = a n^{{1}/{3}} \, ,
\label{delta}
\end{equation}
with $n\!\equiv\! N/V$,
which completely characterizes the properties of
$\delta\!\ll\! 1$.

The corresponding condensate wave function
and the quasiparticle eigenfunctions are
the plane waves:
\begin{equation}
\Psi({\bf r})=\sqrt{n_{0}} \, ,
\label{Psi-k}
\end{equation}
\begin{equation}
\hat{u}_{\bf k}({\bf r})=
\frac{1}{\sqrt{V}}{\rm e}^{i{\bf k}\cdot{\bf r}}
\left[
\begin{array}{cc}
\vspace{1mm}
u_{\bf k} & -v_{\bf k}
\\
-v_{\bf k} & u_{\bf k}
\end{array}
\right] ,
\label{uHat-k}
\end{equation}
with $n_{0}\! \equiv\! {N}_{0}/{V}$ denoting the condensate density.
It should be pointed out that the definition of 
$\hat{u}_{\bf k}({\bf r})$ above
is slightly different from Eq.\ (\ref{uHat}).
Indeed, $\hat{u}_{\bf k}({\bf r})$ is composed of 
$\vec{u}_{\bf k}({\bf r})$ and 
$-\hat{\tau}_{3}\vec{u}_{-{\bf k}}^{\,*}({\bf r})$
with the common spatial dependence 
${\rm e}^{i{\bf k}\cdot{\bf r}}$, which is more convenient
for the homogeneous case.
The coefficients $u_{\bf k}$ and $v_{\bf k}$
are chosen real; they clearly have the symmetry:
$u_{-{\bf k}}\!=\!u_{\bf k}$
and $v_{-{\bf k}}\!=\!v_{\bf k}$.

Accordingly, Eq.\ (\ref{hatSigma})
can be expanded as
\begin{equation}
\hat{\Sigma}({\bf r},{\bf r}')=
\frac{1}{V}\sum_{\bf k}{\rm e}^{i{\bf k}\cdot({\bf r}-{\bf r}')}
\left[
\begin{array}{cc}
\vspace{1mm}
\Sigma & \Delta
\\
-\Delta & -\Sigma
\end{array}
\right] ,
\label{hatSigma-k}
\end{equation}
where the coefficients $\Sigma$ and $\Delta$
have no ${\bf k}$ dependence for 
the contact interaction.
The Hartree-Fock energy $\Sigma$ is obtained 
from Eqs.\ (\ref{Sigma}), (\ref{N}), and (\ref{V}) as
\begin{equation}
\Sigma= \frac{8\pi\hbar^{2} n a}{m} \, .
\label{Sigma-k}
\end{equation}
It merely shifts the
chemical potential and does not play an important
role in the thermodynamics
with constant density.
Substitution of Eqs.\ (\ref{Psi-k}) and (\ref{hatSigma-k})
into Eq.\ (\ref{GP}) yields
the Hugenholtz-Pines relation:
\begin{equation}
\mu_{\rm s}=\Sigma-\Delta \, .
\label{H-P}
\end{equation}
Hence the excitation spectrum has no energy gap in our 
mean-field theory.

The quasiparticle eigenstates are obtained by
substituting Eqs.\ (\ref{uHat-k}) and (\ref{hatSigma-k})
into Eqs.\ (\ref{hatH}) and (\ref{BdG})
and diagonalizing the resulting $2\!\times\!2$
matrix in ${\bf k}$ space. 
We thereby arrive at the well-known expressions:\cite{Fetter72}
\begin{subequations}
\begin{equation}
E_{\bf k}\!=\!\sqrt{\epsilon_{\bf k}(\epsilon_{\bf k}\!+\! 2\Delta)}
\, ,
\label{E_k}
\end{equation}
\begin{equation}
u_{\bf k}\!=\!\sqrt{\frac{\xi_{\bf k}\!+\! E_{\bf k}}{2E_{\bf k}}} \, ,
\hspace{5mm} 
v_{\bf k}\!=\!\sqrt{\frac{\xi_{\bf k}\!-\! E_{\bf k}}{2E_{\bf k}}}\, ,
\end{equation}
\end{subequations}
with $\epsilon_{\bf k}\!=\! \hbar^{2}k^{2}/2m$ and
$\xi_{\bf k}\!=\!\epsilon_{\bf k}\!+\!\Delta$.
Putting them back into Eqs.\ (\ref{Delta}) and (\ref{N}) 
with Eq.\ (\ref{GF}), we obtain 
\begin{equation}
n = n_{0} + A\left(
\int_{0}^{\infty} \frac{\xi}{E}\,
\frac{\varepsilon^{1/2}}{{\rm e}^{E/T}-1}\, d\varepsilon
+\frac{\sqrt{2}}{3}\Delta^{3/2}\right),
\label{n-e}
\end{equation}
\begin{equation}
\Delta = \frac{4\pi \hbar^{2} a}{m} \left[n_{0}
+A\left(\int_{0}^{\infty}\frac{\Delta}{E}
\frac{\varepsilon^{1/2}}{{\rm e}^{E/T}-1}\, d\varepsilon
+ \Delta\varepsilon_{\rm c}^{1/2}
\right)
\right] ,
\label{Delta-e}
\end{equation}
where $A\!\equiv\! m^{3/2}/\sqrt{2}\pi^{2}\hbar^{3}$ is 
a numerical constant, and we have introduced an energy cutoff 
$\varepsilon_{\rm c}$ in Eq.\ (\ref{Delta-e}) to remove an ultraviolet
divergence inherent in the contact interaction.
Equations (\ref{n-e}) and (\ref{Delta-e}) forms 
coupled self-consistent equations
for $n_{0}$ and $\Delta$, which completely
determines the thermodynamic equilibrium
at a given temperature.

Once $n_{0}$ and $\Delta$ are obtained by 
Eqs.\ (\ref{n-e}) and (\ref{Delta-e}), 
we can calculate equilibrium thermodynamic properties 
such as 
the pressure $p_{\rm s}$, the entropy $S_{\rm s}$, and 
the superfluid density $\rho_{\rm s}$.
Pressure $p_{\rm s}\!=\! -\Omega_{\rm eq}/{V}$ is transformed from Eq.\ 
(\ref{Omega-eq}) into
\begin{eqnarray}
&&\hspace{-5mm}
p_{\rm s} =  \frac{\Sigma }{2}n
\!-\! \frac{\Delta }{2}n_{0}
+A\int_{0}^{\infty}\left(\!
\frac{2\varepsilon\xi}{3E}\!-\!\frac{\Delta^{2}}{2E}\!\right)
\frac{\varepsilon^{1/2}}{{\rm e}^{E/T}-1} \, d\varepsilon 
\nonumber \\
&&\hspace{3mm}
-\frac{A}{15\sqrt{2}}\Delta^{5/2}.
\label{p-e}
\end{eqnarray}
Equation (\ref{S}) for $S_{\rm s}$ now reads
\begin{equation}
\frac{S_{\rm s}}{N} = \frac{A}{nT}
\!\int_{0}^{\infty}\!
\frac{5\xi\!+\! 3 \Delta}{3E}
\frac{\varepsilon^{3/2}}{{\rm e}^{E/T}-1}\,
d\varepsilon \, .
\label{S-e}
\end{equation}
The expression of $\rho_{\rm s}$ is given 
by\cite{Fetter72}
\begin{equation}
\rho_{\rm s}=mn\!\left[1-\frac{2A}{3nT}\int_{0}^{\infty}
\frac{\varepsilon^{3/2}{\rm e}^{E/T}}
{({\rm e}^{E/T}-1)^{2}}\, d\varepsilon \right] .
\label{rhoS1}
\end{equation}
The constant $A$ in Eqs.\ (\ref{n-e})-(\ref{rhoS1}) 
under constant density
may be written alternatively in terms of 
the transition temperature of the ideal Bose gas
$T_{0}\!=\!({2\pi \hbar^{2}}/{m})
\bigl[n/\zeta(\frac{3}{2})\bigr]^{{2}/{3}}$ as
\begin{equation}
A=\frac{ 2n}{\sqrt{\pi}\zeta(\frac{3}{2})T_{0}^{{3}/{2}}}\, ,
\end{equation}
with $\zeta(\frac{3}{2})\!=\!2.61$ denoting the Riemann
$\zeta$ function.
It is convenient to choose $T_{0}$ as the unit of energy.
Then Eq.\ (\ref{Delta-e}), for example, reads
\begin{eqnarray}
&&\hspace{-5mm}\left[1-\frac{4\epsilon_{c}^{{1}/{2}}\delta}
{\sqrt{\pi}\zeta(\frac{3}{2})^{{1}/{3}}}
\right] \! \Delta=2\zeta({\textstyle\frac{3}{2}})^{{2}/{3}}
\delta\,
\biggl[\,\frac{n_{0}}{n}
\nonumber \\
&&\hspace{10mm}
+\frac{2}{\sqrt{\pi}\zeta(\frac{3}{2})}
\int_{0}^{\infty}\frac{\Delta}{E}
\frac{\varepsilon^{{1}/{2}}}{{\rm e}^{E/T}-1}\, d\varepsilon
\,\biggr]  .
\label{Delta-e2}
\end{eqnarray}
We choose the cutoff $\epsilon_{c}$ so that 
$1\!\ll\!\epsilon_{c}\!\ll\! 
0.3\delta^{-2}$ is satisfied
and neglect the cutoff-dependent term 
in the following.
This does not affect the results qualitatively,
and even quantitatively to the leading order in $\delta$.

Numerical calculations of Eqs.\ (\ref{n-e})-(\ref{rhoS1}) 
can be performed easily to
clarify temperature dependences of the basic thermodynamic
quantities in the condensed phase.
A change of variable $\varepsilon\!=\!
2\Delta \sinh^{2} t$ is found to improve convergence of the integrations
including ${\rm e}^{E/T}$.
The corresponding equations for the normal state
are obtained from Eqs.\ (\ref{n-e}), (\ref{p-e}), and (\ref{S-e})
by setting $\Delta\!=\! n_{0}\!=\! 0$ and $E\!=\!\xi\!=\!
\varepsilon\!+\!\Sigma\!-\!\mu_{\rm n}$, which yield
$\mu_{\rm n}$, $p_{\rm n}$, and $S_{\rm n}$ as a function of $T$.
The transition temperature $T_{c}$
is then determined by the thermodynamic condition
$F_{\rm s}(T_{c},V,N)\!=\!F_{\rm n}(T_{c},V,N)$ appropriate
under constant volume.
The specific heat $C$ may be calculated
from the results on $S$ by numerical differentiations.

\subsection{Properties near $T_{c}$}
\label{subsec:nearTc}

We now investigate the properties near $T_{c}$ $(\sim\! T_{0}\!=\! 1)$ 
in more detail.
To this end, we make use of the following
asymptotic expansions for the integrals
in Eqs.\ (\ref{n-e})-(\ref{p-e}), respectively:
\begin{subequations}
\label{Is}
\begin{equation}
\frac{2}{\sqrt{\pi}\zeta(\frac{3}{2})}\int_{0}^{\infty}
\frac{\xi}{E}\,
\frac{\varepsilon^{1/2}}{{\rm e}^{E/T}-1}\,
d\varepsilon = T^{3/2}(1-b_{1}x^{1/2}) \, ,
\label{In}
\end{equation}
\begin{equation}
\frac{2}{\sqrt{\pi}\zeta(\frac{3}{2})}\int_{0}^{\infty}
\frac{\Delta}{E}\,
\frac{\varepsilon^{1/2}}{{\rm e}^{E/T}-1}\,
d\varepsilon = T^{3/2} b_{1} x^{1/2} \, ,
\label{Id}
\end{equation}
\begin{eqnarray}
&&\hspace{-3mm}
\frac{2}{\sqrt{\pi}\zeta(\frac{5}{2})}\int_{0}^{\infty}
\left(\frac{2\varepsilon\xi}{3E}-
\frac{\Delta^{2}}{2E}
\right)\frac{\varepsilon^{1/2}}{{\rm e}^{E/T}-1}\,
d\varepsilon  
\nonumber \\
&& \hspace{-6mm} = T^{5/2}
(1-b_{1}''x+b_{2}''x^{3/2}) \, ,
\label{Ip}
\end{eqnarray}
\end{subequations}
with $x\!\equiv\! \Delta/T$, 
$b_{1}\!\equiv\! \sqrt{2\pi}/\zeta(\frac{3}{2})\!=\! 0.960$,
$b_{1}''\!\equiv\!
\zeta(\frac{3}{2})/\zeta(\frac{5}{2})\!=\! 1.95$, and 
$b_{2}''\!\equiv\! 5\sqrt{2\pi}/6\zeta(\frac{5}{2})\!=\!
1.56$.
Each expression of Eq.\ (\ref{Is}) has been obtained 
as follows: (i) Expand the integrand in terms of $\Delta$ 
up to the order where the integral converges
to obtain the leading terms analytic in $x$.
(ii) Subtract the leading contribution from the original integral
to express the residual term as an integral.
(iii) Estimate the higher-order integral
by expanding the exponentials in the denominators up to the first order.
For example, analytic terms are absent in Eq.\ (\ref{Id}),
and the integral has been estimated by approximating
${\rm e}^{E/T}\!-\!1\!\approx\! E/T$ in the denominator.
Those residual integrals generally yield terms
non-analytic in $x$.
This non-analyticity in the expansion near $T_{c}$ 
is inherent in Bose systems
stemming from the divergence of the Bose distribution 
function at zero energy.
This is in marked contrast to
the weak-coupling theory of superconductivity 
where the gap equation near $T_{c}$ is analytic in 
$x^{2}\!\equiv\! (\Delta/T)^{2}$
to make the Landau-Ginzburg expansion possible.\cite{AGD63}
Thus, unlike the Ginzburg-Landau equations for superconductors,\cite{GL50} 
the Ginzburg-Pitaevskii equation\cite{GP58}
for the condensed Bose systems
cannot be justified microscopically.
It is this non-analyticity which turns the mean-field transition
into first order and also brings various anomalies 
in the thermodynamic properties near $T_{c}$
which cannot be described by
the Landau theory of second-order transition.\cite{LandauV}

Substituting Eqs.\ (\ref{In}) and (\ref{Id}) into
Eqs.\ (\ref{n-e}) and (\ref{Delta-e}), respectively,
we obtain
\begin{subequations}
\label{n0}
\begin{equation}
\frac{n_{0}}{n}= 1- T^{3/2}(1-b_{1}y+b_{2}y^{2}) 
\, ,
\label{n0-1}
\end{equation}
\begin{equation}
\frac{n_{0}}{n}=\frac{1+b_{2}'c T^{1/2}\delta}{c\delta}
T y\!\left(y-\frac{b_{1}c T^{1/2}\delta}
{1\!+\!b_{2}'c T^{1/2}\delta}\right) ,
\label{n0-2}
\end{equation}
\end{subequations}
with $y\!\equiv\! x^{1/2}\!=\!
(\Delta/T)^{1/2}$,
$b_{2}\!=\! 0.559$, $b_{2}'\!=\! 1.12$, and
$c\!=\! 2\zeta(\frac{3}{2})^{{2}/{3}}\!=\! 3.79$.
Here we have used the results of Eqs.\ (\ref{In}) 
and (\ref{Id}) including higher-order terms $b_{2}x$ and
$-b_{2}'x$, respectively, with
$b_{2}$ and $b_{2}'$ estimated numerically.\cite{comment-exp}
Let us subtract Eq.\ (\ref{n0-2}) from Eq.\ (\ref{n0-1}).
Solving the resulting equation, 
we obtain $y$ as a function of $T$ as
\begin{eqnarray}
&&\hspace{-10mm}
y(T)=\frac{b_{1}c T^{1/2}\delta}
{1\!+\! (b_{2}\!+\!b_{2}')c T^{1/2}\delta}\biggl\{
1\!+\! \biggl[ 
1\!-\! \frac{T^{-\frac{1}{2}}\!-\! T^{-2}}{b_{1}^{2}c\delta}
\nonumber \\
&&\hspace{2mm} \times
[1\!+\! (b_{2}\!+\!b_{2}')c T^{1/2}\delta]\,\biggr]^{1/2}
\biggr\} \, .
\label{y(T)}
\end{eqnarray}
The transition temperature $T_{c}$ may be estimated by setting 
$n_{0}\!=\! 0$ in Eq.\ (\ref{n0-2}), which
yields
\begin{equation}
y(T_{c})-\frac{b_{1}c T_{c}^{1/2}\delta}
{1\!+\!b_{2}'c T_{c}^{1/2}\delta}=0 \, .
\label{y(T_c)}
\end{equation}
Equation (\ref{y(T_c)}) with Eq.\ (\ref{y(T)}) 
can be solved by expanding $T_{c}$
in powers of $\delta$ as
\begin{eqnarray}
&&\hspace{-8mm}
T_{c}=1+\frac{8\pi}{3\zeta(\frac{3}{2})^{4/3}}
\delta +
\frac{8\pi}{3\zeta(\frac{3}{2})^{4/3}} \! \left[
\frac{14\pi}{3\zeta(\frac{3}{2})^{4/3}}
\!-\! (b_{2}\!+\! b_{2}')c\right]\!\delta^{2}
\nonumber \\
&&\hspace{-3.5mm}
=1+2.33\delta+\cdots \, .
\label{Tc}
\end{eqnarray}
Thus, $T_{c}$ initially increases linearly with $\delta$.
The numerical coefficient $2.33$ agrees with 
the analytic result by Baym {\em et al}.\ \cite{Baym00}
as well as the numerical one by 
Holzmann and Krauth.\cite{Holzmann99}
It is worth pointing out that this expression
makes the coefficients of $\delta^{0}$ and $\delta$
in the large square bracket of Eq.\ (\ref{y(T)}) vanish at $T_{c}$.
Substituting Eq.\ (\ref{Tc}) into Eq.\ (\ref{y(T_c)}) and
noting $y\!\equiv\!(\Delta/T)^{1/2}$,
we obtain
\begin{equation}
\Delta(T_{c})
=\frac{8\pi}{\zeta(\frac{3}{2})^{2/3}}\delta^{2} = 
13.3\delta^{2} \, ,
\label{Dc}
\end{equation}
to the leading order in $\delta$.

A comment is necessary on Eq.\ (\ref{Tc}) at this stage.
The expression (\ref{Tc}) has been obtained as the highest temperature
of the non-trivial solutions to Eq.\ (\ref{n0}).
However, the corresponding superfluid transition is first-order.
It hence follows that, under the present conditions of 
fixed $V$ and $N$, the transition temperature should have been determined 
alternatively by the requirement that the free energy $F(T,V,N)$ 
be equal between the normal and superfluid phases.
It will be shown below Eq.\ (\ref{Dp}), however, that Eq.\ (\ref{Tc})
indeed satisfies the requirement up to the order $\delta^{2}$.

We next consider the region below $T_{c}$
and expand $y(T) -y(T_{c})$
in powers of $1\!-\! T/T_{c}$.
As noted below Eq.\ (\ref{Tc}),
the leading term at $T_{c}$ in the large square bracket of Eq.\ (\ref{y(T)})
is proportional to $\delta^{2}$.
The proportionality constant is found to be $(b_{2}c)^{2}$
from the equation of order $\delta^{3}$ in the expansion of
Eq.\ (\ref{Tc}).
Due to this smallness of the constant term at $T\!=\! T_{c}$, 
the temperature-dependent contribution
$\propto (1\!-\! T/T_{c})$ in the square bracket may not
be placed outside.
Indeed, we obtain
\begin{eqnarray}
&&\hspace{-11mm}
y(T)-y(T_{c})
\nonumber \\
&&\hspace{-11mm}
=-b_{1}b_{2}c^{2}\delta^{2}
+\left[(b_{1}b_{2}c^{2}\delta^{2})^{2}
+\frac{3}{2}c\delta (1\!-\! T/T_{c})\right]^{1/2}
\nonumber \\
&&\hspace{-11mm}
\approx \left\{\!\!
\begin{array}{ll}
\vspace{2mm}
0.369\delta^{-1}(1\!-\! T/T_{c})
& 
: 1\!-\! T/T_{c}\!\ll\! 10.5\delta^{3}
\\
\sqrt{3}\zeta({\textstyle\frac{3}{2}})^{1/3}
\delta^{1/2}(1\!-\! T/T_{c})^{1/2} 
& : 1\!-\! T/T_{c}\!\gg\!10.5\delta^{3}
\end{array}
\right.
 \!\!\!  .
 \nonumber \\
\label{yyc}
\end{eqnarray}
Noting $y\!=\!(\Delta/T)^{1/2}$ and Eq.\ (\ref{Dc}),
we may approximate $\Delta$ just below $T_{c}$ as
\begin{eqnarray}
&&\hspace{-14mm}
\Delta (T) -\Delta(T_{c})
\nonumber \\
&&\hspace{-14mm}
=\left\{\!\!
\begin{array}{ll}
\vspace{2mm}
2.69(1\!-\! T/T_{c})
& 
: 1\!-\! T/T_{c}\!\ll\! 10.5\delta^{3}
\\
\sqrt{96\pi}\,
\delta^{3/2}(1\!-\! T/T_{c})^{1/2}
& 
: 1\!-\! T/T_{c}\!\gg\!10.5\delta^{3}
\end{array}
\right.
 \!\!\! .
\label{DeltaTTc}
\end{eqnarray}
Let us substitute Eqs.\ (\ref{Dc}) and 
(\ref{yyc}) into Eq.\ (\ref{n0-2}).
We then find that $n_{0}(T)$ grows as
\begin{equation}
\frac{n_{0}}{n} = 
\left\{\!\!
\begin{array}{ll}
\vspace{2mm}
0.354\delta^{-1}(1\!-\! T/T_{c})
\! & \!
: 1\!-\! T/T_{c}\!\ll\! 10.5\delta^{3}
\\
\displaystyle
\frac{\sqrt{6\pi}}
{\zeta(\frac{3}{2})^{2/3}} 
\delta^{1/2}(1\!-\! T/T_c)^{1/2}\! & \! 
: 1\!-\! T/T_{c}\!\gg\!10.5\delta^{3}
\end{array}
\right.\!\!
.
\label{n0TTc}
\end{equation}
Equations (\ref{Tc}), (\ref{Dc}), (\ref{DeltaTTc}), and (\ref{n0TTc})
are the main results obtained from Eq.\ (\ref{n0}).

We turn our attention to other thermodynamic quantities.
The chemical potential below $T_{c}$ satisfies the Hugenholtz-Pines
relation $\mu_{\rm s}(T)\!=\!\Sigma\!-\!\Delta(T)$ and 
directly reflects the singularity of $\Delta(T)$
near $T_c$.
On the other hand, $\mu_{\rm n}(T)$
in the normal state of $T\!\gtrsim\!1$ obeys
\begin{equation}
\mu_{\rm n}(T) = \Sigma - \frac{\zeta(\frac{3}{2})^{2}}{4\pi }
(T^{3/2}\!-\! 1)^{2} \, .
\label{mu_n}
\end{equation}
This can be shown from ${N}\!=\! \sum_{\bf k}
[{\rm e}^{(\varepsilon_{\bf k}+\Sigma-\mu_{\rm n})/T}\!-\! 1]^{-1}$
with the procedure of deriving the expansions of
Eq.\ (\ref{Is}).
Using Eqs.\ (\ref{Tc}) and (\ref{Dc}), we obtain the discontinuity
$\Delta\mu(T_{c})\!\equiv\! \mu_{\rm s}(T_{c})\!-\! \mu_{\rm n}(T_{c})$
as
\begin{equation}
\Delta\mu(T_{c})=-
\frac{4\pi}{\zeta(\frac{3}{2})^{2/3}}\delta^{2}  =-6.63 \delta^{2}\, .
\label{dMu}
\end{equation}

Next, with the procedure of deriving Eq.\ (\ref{Is}), 
the entropy just below $T_{c}$ is calculated from
Eq.\ (\ref{S-e}) as
\begin{equation}
\frac{S_{\rm s}}{N}=\frac{5\zeta(\frac{5}{2})}{2\zeta(\frac{3}{2})}
T^{3/2}-\frac{3}{2}T^{1/2}\Delta \, .
\label{STTc}
\end{equation}
The corresponding normal-state expression for $T\!\gtrsim\!1$
is given by
\begin{equation}
\frac{S_{\rm n}}{N}=\frac{5\zeta(\frac{5}{2})}{2\zeta(\frac{3}{2})}
T^{3/2}-\frac{3\zeta(\frac{3}{2})^{2}}{8\pi }
(T^{3/2}\!-\! 1)^{2} \, .
\end{equation}
Hence the discontinuity
$\Delta S(T_{c})\!\equiv\! S_{\rm s}(T_{c})\!-\! S_{\rm n}(T_{c})$
is found by using Eqs.\ (\ref{Tc}) and (\ref{Dc}) as
\begin{equation}
\Delta S(T_{c})=-
\frac{6\pi}{\zeta(\frac{3}{2})^{2/3}}\delta^{2}  
=-9.94 \delta^{2} \, .
\end{equation}
It follows from Eqs.\ (\ref{DeltaTTc}) and (\ref{STTc}) that
the specific heat $C_{\rm s}=T(\partial S_{\rm s}/\partial T)$ just below
$T_{c}$ is given to the leading order by
\begin{eqnarray}
&&\hspace{-10mm}
\frac{C_{\rm s}}{N}=\frac{15\zeta(\frac{5}{2})}{4\zeta(\frac{3}{2})}
T^{3/2} 
-\frac{3}{2}T^{3/2}\frac{d\Delta}{dT}
\nonumber \\
&&\hspace{-10mm}
=\frac{C_{\rm n}}{N}+\left\{\!\!
\begin{array}{ll}
\vspace{2mm}
4.04
& 
: 1\!-\! T/T_{c}\!\ll\! 10.5\delta^{3}
\\
\displaystyle
\frac{3\sqrt{6\pi}
\delta^{3/2}}{(1-T/T_c)^{1/2}} & 
: 1\!-\! T/T_{c}\!\gg\!10.5\delta^{3}
\end{array}
\right.
 \! .
\label{CTTc}
\end{eqnarray}
It displays a divergent behavior
$C_{\rm s}/N\! \sim \!13.0 \delta^{3/2}(1\!-\! T/T_c)^{-\frac{1}{2}}$
for $10.5\delta^{3}\!\ll\!1\!-\! T/T_{c}\!\ll\! 1$.

Substituting Eq.\ (\ref{Ip})
into Eq.\ (\ref{p-e}),
$p_{\rm s}$ just below $T_c$ 
is obtained  as
\begin{equation}
\frac{p_{\rm s}}{n} = \frac{\Sigma }{2}
- \frac{\Delta }{2}\frac{n_{0}}{n}
+ 
\frac{\zeta(\frac{5}{2})}{\zeta(\frac{3}{2})}T^{5/2}
-T^{3/2}\Delta \, .
\label{pTTc}
\end{equation}
The normal-state pressure of $T\!\gtrsim\! 1$ can be calculated
similarly to be
\begin{equation}
\frac{p_{\rm n}}{n} = \frac{\Sigma }{2}
+ \frac{\zeta(\frac{5}{2})}{\zeta(\frac{3}{2})}T^{5/2}
-\frac{\zeta(\frac{3}{2})^{2}}{4\pi}
(T^{3/2}-1)^{2} \, .
\label{p_n}
\end{equation}
Hence the discontinuity
$\Delta p(T_{c})\!\equiv\! p_{\rm s}(T_{c})\!-\! p_{\rm n}(T_{c})$
is obtained as
\begin{equation}
\frac{\Delta p(T_{c})}{n}=-
\frac{4\pi}{\zeta(\frac{3}{2})^{2/3}}\delta^{2} =-6.63 \delta^{2}
\, .
\label{Dp}
\end{equation}
It follows from Eqs.\ (\ref{dMu}) and (\ref{Dp}) as well as 
$F(T,V,N)\!=\! -p V\!+\!\mu N$
that $F_{\rm s}(T_{c},V,N)\!=\! F_{\rm n}(T_{c},V,N)$ is satisfied
for the expression (\ref{Tc}) up to the order $\delta^{2}$.
Thus, Eq.\ (\ref{Tc}) is indeed a correct expression of 
$T_{c}$ of order $\delta$.

\begin{figure}[t]
\begin{center}
\includegraphics[width=7cm]{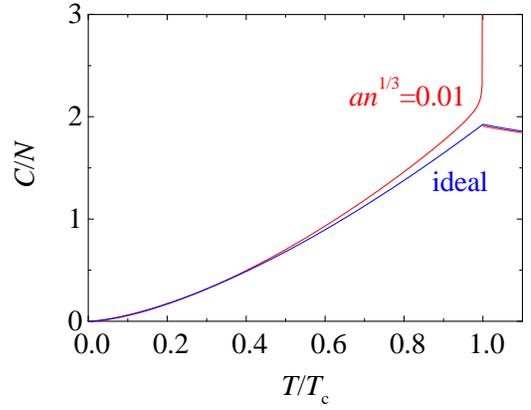}%
\end{center}
\caption{ Specific heat $C/N$ as a function of $T/T_c$
for $an^{1/3}\!=\! 0.01$ shown in comparison with the ideal gas result.
}
\label{fig:2}
\end{figure}

We finally consider the superfluid density near $T_{c}$.
With the procedure of deriving Eq.\ (\ref{Is}), 
Eq.\ (\ref{rhoS1}) is transformed into
\begin{equation}
\frac{\rho_{\rm s}}{mn}=1-
T^{3/2}
+\frac{4\sqrt{2\pi}T\Delta^{1/2}}
{3\zeta(\frac{3}{2})} 
\, .
\label{rhoSTTc}
\end{equation}
Substituting Eqs.\ (\ref{Tc}) and (\ref{Dc})
into Eq.\ (\ref{rhoSTTc}),
we obtain
\begin{equation}
\frac{\Delta\rho_{\rm s}(T_{c})}{mn}=
\frac{4\pi}
{3\zeta(\frac{3}{2})^{4/3}} \delta = 1.16\delta
\, .
\label{DrhoS}
\end{equation}

\subsection{Properties at $T=0$}

Physical properties at $T\!=\!0$ are easily found from Eqs.\
(\ref{H-P}) and
(\ref{n-e})-(\ref{rhoS1}) 
by dropping terms with the Bose distribution function.
The eading-order expressions are summarized as follows:
\begin{subequations}
\begin{equation}
\frac{n_{0}(0)}{n} = 1-\frac{8}{3\sqrt{\pi}} \delta^{3/2}
= 1-1.50 \delta^{3/2}\, ,
\label{n0T0}
\end{equation}
\begin{equation}
{\mu_{\rm s}(0)} ={\Delta(0)}=2\frac{p_{\rm s}(0)}{n}
= 2\zeta({\textstyle\frac{3}{2}})^{2/3} \delta= 3.79 \delta\, .
\label{muT0}
\end{equation}
\end{subequations}
\begin{figure}[t]
\begin{center}
\includegraphics[width=6.5cm]{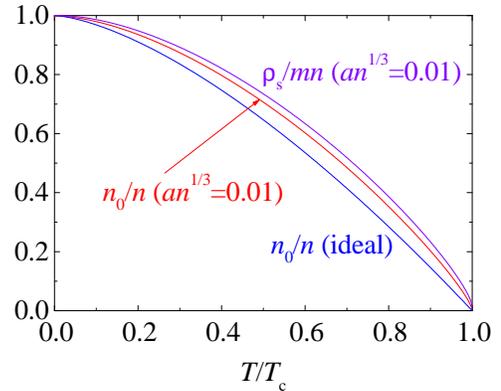}%
\end{center}
\caption{ Normalized condensate density $n_{0}/n$
and superfluid density $\rho_{\rm s}/nm$ as a function
of $T/T_c$ for $an^{1/3}\!=\! 0.01$.
The ideal gas result 
($n_{0}/n\!=\!\rho_{\rm s}/nm$) is also plotted for comparison.
}
\label{fig:3}
\end{figure} 
\begin{figure}[t]
\begin{center}
\includegraphics[width=7.5cm]{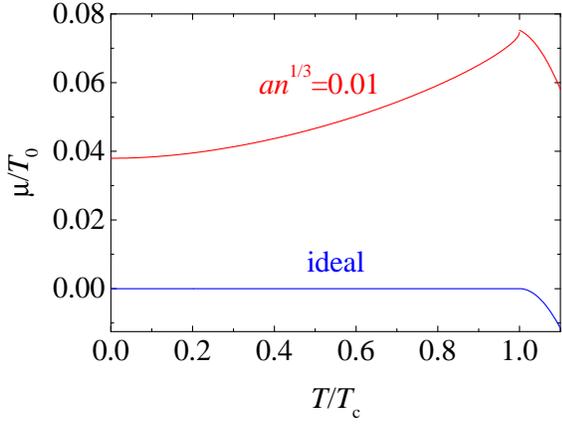}%
\end{center}
\caption{ Temperature dependence of the chemical potential $\mu$
for $an^{1/3}\!=\! 0.01$ shown in comparison with the ideal gas result.
}
\label{fig:4}
\end{figure}

\subsection{Numerical Results}

Figure \ref{fig:2} displays
the specific heat $C/N$
as a function of the renormalized temperature
$T/T_{c}$ for $an^{1/3}\!=\! 0.01$.
The ideal-gas result ($a\!=\! 0; T_c\!=\! T_0$)
is also plotted for comparison.
The curve for $an^{1/3}\!=\! 0.01$
is proportional to $T^{3}$ at very low temperatures
due to $E_{\bf k}\!\propto\! k$
for $k\!\rightarrow\! 0$.
As the temperature is increased,
the curve starts to deviate upward from the ideal-gas result
to show the limiting behavior (\ref{CTTc})
for $T\!\lesssim\! T_{c}$.

Figure \ref{fig:3} plots $n_0/n$ and $\rho_{\rm s}/nm$
as a function of $T/T_c$ for $an^{1/3}\!=\! 0.01$
and $0$.
We observe that the finite interaction causes an enhancement
of $n_0/n$ and $\rho_{\rm s}/nm$ over the 
ideal-gas value at all temperatures.
The enhancement is larger for $\rho_{\rm s}/nm$ than
$n_0/n$. The quantity $n_0/n$ obeys Eq.\ (\ref{n0TTc}) near $T_{c}$ and
develops continuously from $0$ even with a
finite interaction,
whereas $\rho_{\rm s}/nm$ starts from a positive value at
$T_{c}$ as Eq.\ (\ref{DrhoS}).

Figure \ref{fig:4} shows 
the chemical potential as a function of
$T/T_c$ for $an^{1/3}\!=\! 0.01$ and $0$.
With a finite interaction,
the chemical potential displays a characteristic peak at $T_{c}$
with a discontinuity,
as shown explicitly by Eq.\ (\ref{dMu}).
It eventually approaches the value of Eq.\ (\ref{muT0})
as the temperature is lowered.

\subsection{The HFB and Shohno theories}
\label{subsec:HFB}

We now compare the above predictions of the conserving gapless
mean-field theory with those of the HFB\cite{GA59,Luban62,RS69,Griffin96} 
and Shohno\cite{Shohno64,RS69} theories.
Although extensive theoretical studies have been carried 
out based on the latter theories, their 
predictions on the thermodynamic quantities seem 
not to have been clarified completely, especially near $T_{c}$.
We hence study them in detail with a special focus
on the region $T\!\lesssim\! T_{c}$.
This will also help us to understand common features
in the mean-field theories of the homogenous weakly interacting
Bose gas.
The transition temperature will be determined below
in the same way as Eq.\ (\ref{Tc}); the two comments below
Eqs.\ (\ref{Dc}) and (\ref{Dp}) also apply
to the present cases.

The equations to determine the equilibrium in the HFB theory 
are formally obtained from Eqs.\ (\ref{n-e}) and (\ref{Delta-e})
by a couple of modifications: (i) a sign change for the second
term in the square bracket of Eq.\ (\ref{Delta-e});
(ii) the expressions of $\xi_{\bf k}$ 
and $E_{\bf k}$.
Indeed, they are given by
\begin{subequations}
\label{DN-HFB}
\begin{equation}
n = n_{0} + \frac{1}{V}\sum_{\bf k} \left(\frac{\xi_{\bf k}}{E_{\bf k}}\,
\frac{1}{{\rm e}^{E_{\bf k}/T}-1}+
\frac{\xi_{\bf k}\!-\! E_{\bf k}}{2E_{\bf k}}
\right) ,
\label{DN0-HFB}
\end{equation}
\begin{equation}
\Delta = \frac{4\pi \hbar^{2} a}{m} n_{0}-\Gamma  \, ,
\label{DN1-HFB}
\end{equation}
\end{subequations}
where $\Gamma$, $\xi_{\bf k}$, and $E_{\bf k}$ are defined by
\begin{equation}
\Gamma\equiv 
\frac{4\pi \hbar^{2} a}{mV}\sum_{\bf k}
\left(\frac{\Delta}{E_{\bf k}} \frac{1}{{\rm e}^{E_{\bf k}/T}-1}
+\frac{\Delta}{2E_{\bf k}}\right) ,
\label{Gamma}
\end{equation}
\begin{equation}
\xi_{\bf k}\equiv \varepsilon_{\bf k}\!+\!\Delta\!+\! 2\Gamma \, ,
\hspace{2mm}
E_{\bf k}\equiv \sqrt{(\varepsilon_{\bf k}\!+\!2\Gamma)
(\varepsilon_{\bf k}\!+\!2\Delta\!+\! 2\Gamma)} \, ,
\label{E-HFB}
\end{equation}
with $\varepsilon_{\bf k}\!\equiv\! \hbar^{2}k^{2}/2m$.
The Shohno theory is obtained from Eqs.\
(\ref{DN-HFB}) and (\ref{E-HFB}) by setting $\Gamma\!=\! 0$.

Near $T\!\sim\!T_{0}\!=\! 1$ where 
$\Delta$ and $\Gamma$ are much smaller than
$T_{0}$, Eq.\ (\ref{DN-HFB}) is approximated 
to the leading order by 
\begin{subequations}
\label{n0-HFB}
\begin{equation}
\frac{n_{0}}{n}=1-T^{3/2}\bigl[1-b_{1}
\bigl(\sqrt{\Gamma\!+\!\Delta}
+\sqrt{\Gamma}\,\bigr)\bigr] \, ,
\label{n01-HFB}
\end{equation}
\begin{equation}
\Delta = c\delta \left[\frac{n_{0}}{n}-b_{1}
\bigl(\sqrt{\Gamma\!+\!\Delta}
-\sqrt{\Gamma}\,\bigr)\right] \, ,
\label{n02-HFB}
\end{equation}
\end{subequations}
with $b_{1}\!\equiv\! \sqrt{2\pi}/\zeta(\frac{3}{2})$
and $c\!=\! 2\zeta(\frac{3}{2})^{2/3}$.
These equations correspond to Eq.\ (\ref{n0}) and have
been obtained with the procedure of deriving 
the expansions of Eq.\ (\ref{Is}).
Note that Eq.\ (\ref{n0-HFB}) is also non-analytic in $\Gamma$
and $\Gamma\!+\!\Delta$.
Using $\Delta\!=c\delta (n_{0}/n\!-\! \Gamma)$
from Eq.\ (\ref{DN1-HFB}),
we next rewrite Eq.\ (\ref{n02-HFB}) in terms of $\Gamma$ and
$n_{0}/n$.
We also introduce the dimensionless variables $r$, $s$, and
$t$ through $\sqrt{n_{0}/n}\!=\! b_{1}\sqrt{c\delta}\, r$,
$\sqrt{\Gamma}\!=\! b_{1}c\delta s$, and $T\!=\!
1\!+\! b_{1}^{2}c\delta t$.
Retaining terms of order $\delta$,
Eq.\ (\ref{n0-HFB}) is transformed into
\begin{equation}
r^{2}-r+\frac{3}{2}t=s \, ,
\hspace{5mm}
s = \frac{-1+\sqrt{1+4r}}{2} \, .
\label{n0-HFB2}
\end{equation}
Equation (\ref{n0-HFB2}) yields 
\begin{equation}
f(r,t)\equiv r^{4}-2r^{3}+(2\!+\!3t)r^{2}-(2\!+\!3t)r+\frac{3}{4}t
(2\!+\!3t)=0 \, .
\label{fpt}
\end{equation}
It has a couple of solutions $r_{1}$, $r_{2}\!>\! 0$ for $t\!\approx\! 0$
but no real solution for a large-enough $t$.
Thus, the transition temperature corresponds to the point
where $f\!=\!0$ and $\partial f/\partial r\!=\!0$,
which yields the critical values of $t\!=\! 11/24$ and $r\!=\! 3/4$.
Thus, the transition temperature of the HFB theory
for $\delta\!\rightarrow\! 0$ is given by $T_{c}^{{\rm HFB}}\!=\!
1\!+\!\Delta T_{c}^{{\rm HFB}}$ with
\begin{subequations}
\begin{equation}
\Delta T_{c}^{{\rm HFB}}=
\frac{11\pi}{6\zeta(\frac{3}{2})^{4/3}}\delta 
= 1.60\delta \, .
\end{equation}
The transition is clearly first order.
The discontinuities of various thermodynamic quantities 
at $T_{c}^{{\rm HFB}}$ can be obtained 
with the procedure of Sec.\ \ref{subsec:nearTc} as
\begin{equation}
\frac{n_{0}^{{\rm HFB}}(T_{c})}{n} = 
\frac{9\pi}{4\zeta(\frac{3}{2})^{4/3}}\delta 
= 1.96\delta \, ,
\end{equation}
\begin{equation}
\Delta^{{\rm HFB}}(T_{c}) = 
\frac{5\pi}{2\zeta(\frac{3}{2})^{2/3}}\delta^{2} 
= 4.14\delta^{2} \, ,
\end{equation}
\begin{equation}
\frac{\rho_{\rm s}^{{\rm HFB}}(T_{c})}{mn} = 
\frac{139\pi}{60\zeta(\frac{3}{2})^{4/3}}\delta 
= 2.02\delta \, .
\end{equation}
\end{subequations}
Note that even $n_{0}/n$ is discontinuous at $T_{c}$ in the HFB theory.

The same analysis can be performed for the Shohno
theory.
We thereby obtain
\begin{subequations}
\begin{equation}
\Delta T_{c}^{{\rm Shohno}}=
\frac{2\pi}{3\zeta(\frac{3}{2})^{4/3}}\delta 
= 0.582\delta \, ,
\end{equation}
\begin{equation}
\frac{n_{0}^{{\rm Shohno}}(T_{c})}{n} = 
\frac{\pi}{\zeta(\frac{3}{2})^{4/3}}\delta 
= 0.873\delta \, ,
\end{equation}
\begin{equation}
\Delta^{{\rm Shohno}}(T_{c}) = 
\frac{2\pi}{\zeta(\frac{3}{2})^{2/3}}\delta^{2} 
= 3.31\delta^{2} \, ,
\end{equation}
\begin{equation}
\frac{\rho_{\rm s}^{{\rm Shohno}}(T_{c})}{mn} = 
\frac{5\pi}{3\zeta(\frac{3}{2})^{4/3}}\delta 
= 1.46\delta \, .
\end{equation}
\end{subequations}
Thus, even the Shohno theory predicts an enhancement of
$T_{c}$ over the ideal-gas value $T_{0}$.

As for the overall temperature dependences of various thermodynamic
quantities, the HFB and Shohno theories both yield results qualitatively
similar to those of the previous mean-field theory
except the discontinuity of $n_{0}$ at $T_{c}$.
For example, the specific heat of the HFB and Shohno theories
also display the divergent behavior of $(T_{c}\!-\! T)^{-1/2}$
just below $T_{c}$.
Due to a finite $\Gamma$, the HFB theory generally
predicts larger deviations from the ideal-gas results
than the Shohno theory.

\section{Homogeneous gas under constant density}

We finally study the homogeneous 
weakly interacting Bose gas under the complementary condition
of constant pressure.
Despite its fundamental importance as a subject of
quantum statistical mechanics,
the system seems to have been investigated only
by Reatto and Straley \cite{RS69}
based on the HFB and Shohno theories.
\cite{RS69,Griffin96}
Moreover, their main interest was on superfluid $^{4}$He so that
the properties in the weak-coupling region
remain essentially unexplored.

Let us introduce a dimensionless parameter $\delta_{p}$ by
\begin{equation}
\delta_{p}\!\equiv\! a\left(\! \frac{mp}{2\pi\hbar^{2}}
\!\right)^{\!1/5} \, ,
\end{equation}
which completely characterizes the weak-coupling region of $\delta_{p}
\!\ll\! 1$.
The thermodynamic equilibrium of the condensed phase 
is determined by Eqs.\ (\ref{n-e})-(\ref{p-e}).
They yield the density $n$, the condensate density $n_{0}$,
and the pair potential $\Delta$ as a function of $p$ and $T$.
The prefactor $A$ in Eqs.\ (\ref{n-e})-(\ref{p-e})
may be rewritten conveniently as
\begin{equation}
A = \frac{2n_{\rm c}}{\sqrt{\pi}\zeta(\frac{3}{2})T_{0}^{3/2}}
= \frac{2p}{\sqrt{\pi}\zeta(\frac{5}{2})T_{0}^{5/2}} \, ,
\label{Ap}
\end{equation}
where $n_{c}$ and $T_{0}$ are
the critical density
and the transition temperature
of the ideal Bose gas, respectively.
They are given in terms of $p$ as
$n_{c}\!=\! [mp/2\pi \hbar^{2}\zeta(\frac{5}{2})
]^{3/5}\zeta(\frac{3}{2})$ 
and $T_{0}\!=\! (2\pi \hbar^{2}/m)^{3/5}
[p/\zeta(\frac{5}{2})]^{2/5}$.
The equations for the normal state
are obtained from Eqs.\ (\ref{n-e}), (\ref{p-e}), and (\ref{S-e})
by setting $\Delta\!=\! n_{0}\!=\! 0$ and $E\!=\!\xi\!=\!
\varepsilon\!+\!\Sigma_{\rm n}\!-\!\mu_{\rm n}$; they yield
$n$, $\mu_{\rm n}$, and $S_{\rm n}$ as a function of $p$ and $T$.
The transition temperature $T_{c}$
is then determined by the thermodynamic condition
$\mu_{\rm s}(T_{c},p)\!=\!\mu_{\rm n}(T_{c},p)$ appropriate
under constant pressure.
The specific heat per particle $C_{p}$ may be calculated
from the results on $S$ by numerical differentiations.
In the following, we will adopt the units 
in which $n_{\rm c}\!=\! T_{0}\!=\! 1$ and neglect the cutoff-dependent
term in Eq.\ (\ref{Delta-e}) as was done in the case of 
constant density.

To investigate the properties near $T_{c}$ $(\sim\! 1)$ 
in more detail, 
we make use of the
asymptotic expansions 
of Eq.\ (\ref{Is}).
Equations (\ref{n-e}) and (\ref{Delta-e}) with Eq.\ (\ref{Is})
enable us to express $n_{0}$ and $n$ in terms of
$T$ and $\Delta$ as
$n_{0}\!=\!\Delta/c'\delta_{p}\!-b_{1}T\Delta^{1/2}$
and
$n\!=\! T^{3/2}\!+\!\Delta/c'\delta_{p}
\!-2b_{1}T\Delta^{1/2}$
with $c'\!\equiv\! 2\zeta(\frac{3}{2})/
\zeta(\frac{5}{2})^{1/5}\!=\! 4.93$.
Substituting them into it, 
Eq.\ (\ref{p-e}) becomes a nonlinear equation for 
$\Delta$ as
\begin{eqnarray}
&&\hspace{-5mm}
\frac{b_{1}''}{2c'\delta_{p}}\Delta^{2}-\left(\!
\frac{7b_{1}b_{1}''}{2}\!-\!b_{2}''\!\right)T\Delta^{\!{3}/{2}}
\nonumber \\
&&\hspace{-5mm}
+\left(b_{1}''T^{3/2}
\!+\!4b_{1}^{2}b_{1}''c' T^{2}\delta_{p}\right)\!\Delta
-4b_{1}b_{1}''c'T^{5/2}\delta_{p}
 \Delta^{1/2}
\nonumber \\
&&\hspace{-5mm}+T^{5/2}+b_{1}''c' T^{3}\delta_{p}-1=0 \, .
\label{Delta-eq}
\end{eqnarray}
Thus, the equation is non-analytic in $\Delta$
and completely different from the one in the Landau theory of
second-order transition.\cite{LandauV}
It is convenient to rewrite $T$ in Eq.\ (\ref{Delta-eq}) as
$T\!=\! 1\!-\! \frac{2}{5}b_{1}''c'\delta_{p}\!+\! t\delta_{p}^{2}
\!=\! 1\!-\! 3.84\delta_{p}\!+\! t\delta_{p}^{2}$,
where  $t$ is a parameter appropriate to describe the region $T\!\sim\! 1$.
Then Eq.\ (\ref{Delta-eq})
can be solved analytically by expanding $\Delta$ as
$\Delta\!=\!\Delta_{2}(t)\delta_{p}^{2}\!+\Delta_{3}(t)
\delta_{p}^{3}\!+\!\cdots$.
The expression for $\Delta_{2}(t)$ is obtained as
\begin{subequations}
\label{DS2}
\begin{equation}
\Delta_{2}(t)=\left[2b_{1}c'+\sqrt{(2b_{1}c')^{2}
+\frac{9b_{1}''c^{\prime 2}}{10}-\frac{5}{2b_{1}''}t}\,\right]^{2} \, .
\end{equation}
It hence follows that a superfluid solution exists
for $t\!\leq\! t_{\rm s}\!\equiv\! (2b_{1}''/5)[
(2b_{1}c')^{2}
+9b_{1}''c^{\prime 2}/10]\!=\!103$.
The quantity $\Delta_{3}(t)$ may be expressed
in terms of $t$ and $\Delta_{2}(t)$.
A similar consideration for the normal state
leads to a nonlinear equation for $\Sigma-\mu$.
Expanding 
$\Sigma-\mu= \Sigma_{2}(t)\delta_{p}^{2}+\Sigma_{3}(t)\delta_{p}^{3}
+\cdots$, $\Sigma_{2}(t)$ is found to satisfy
\begin{equation}
\sqrt{\Sigma_{2}(t)}=-\sqrt{2}b_{1}c'+\sqrt{2(b_{1}c')^{2}
-\frac{9b_{1}''c^{\prime 2}}{10}+\frac{5}{2b_{1}''}t}\, .
\end{equation}
\end{subequations}
It has a real solution $\Sigma_{2}$ for 
$t\!\geq\! t_{\rm n}\!\equiv\! 18(b_{1}''c')^{2}/50\!=\! 33.1$.

\begin{figure}[t]
\begin{center}
\includegraphics[width=7cm]{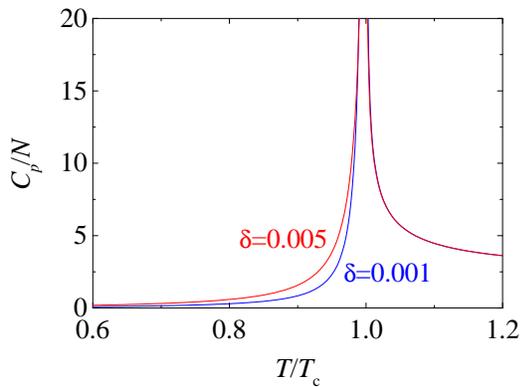}%
\end{center}
\caption{ Specific heat $C_{p}$ as a function of $T/T_c$
for $\delta_{p}\!=\! 0.001$ and $0.005$.
}
\label{fig:5}
\end{figure}
\begin{figure}[t]
\begin{center}
\includegraphics[width=7cm]{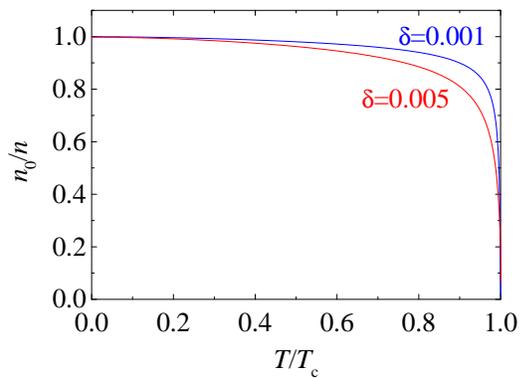}%
\end{center}
\caption{ Normalized condensate density $n_{0}/n$ as a function of $T/T_c$
for $\delta_{p}\!=\! 0.001$ and $0.005$.
}
\label{fig:6}
\end{figure}

From $t_{\rm s}\!>\! t_{\rm n}$, we realize that the superfluid transition
is first order with a metastable region extending over
$t_{\rm n}\!\leq\! t\!\leq\! t_{\rm s}$.
The transition temperature $T_{c}$ can also be expressed as
\begin{equation}
T_{c}= 1-3.84\delta_{p}+t_{c}\delta_{p}^{2} \, .
\label{Tc(p)}
\end{equation}
Thus, $T_{c}$ becomes smaller than $T_{0}\!=\! 1$ 
to the leading order in $\delta_{p}$.
This behavior is opposite to the case of constant volume
where $T_{c}$ increases as a function of 
the dimensionless parameter $an^{1/3}$.\cite{Andersen04}
Indeed, the effect of the weak repulsive interaction 
is completely different between the volume-fixed and pressure-fixed cases.
In the former case, a finite interaction 
suppresses the density fluctuation so that it works favorably 
for the phase coherence over the system, thereby
leading to an enhancement of $T_{c}$. In the latter case, on the other hand,
the interaction lowers the particle density to decrease $T_{c}$.
The value of the prefactor $3.84$ is almost certainly correct,
in contrast to the volume-fixed case where
it seems still controversial.\cite{Andersen04,Yukalov04}

We turn our attention to $t_{c}$ of Eq.\ (\ref{Tc(p)})
which includes all the higher-order contributions.
It is found that $t_{c}$ has a rather large $\delta_{p}$ dependence,
due partly to the large values of the numerical constants 
appearing in the equations, e.g.,
$t_{\rm s}\!=\! 103$.
It decreases from $t_{c}(\delta_{p}
\!\rightarrow\! 0)\!=\! t_{\rm s}\!=\! 103$
down to $98.6$, $96.9$, and $95.7$ at $\delta_{p}\!=\! 5.0\times 10^{-4}$,
$1.0\times 10^{-3}$, and $1.5\times 10^{-3}$, respectively.
The contributions from 
$\Delta_{3}$ and $\Sigma_{3}$ are essential to produce the 
results,
which agree excellently with the exact numerical results
of using Eqs.\ (\ref{n-e})-(\ref{p-e}) for $T\!\sim\! 1$.
Discontinuities in the thermodynamic quantities at $T_{c}$ are
found to be all of the order of $\delta_{p}^{2}$ in accordance 
with the tiny metastable region.

\begin{figure}[t]
\begin{center}
\includegraphics[width=7cm]{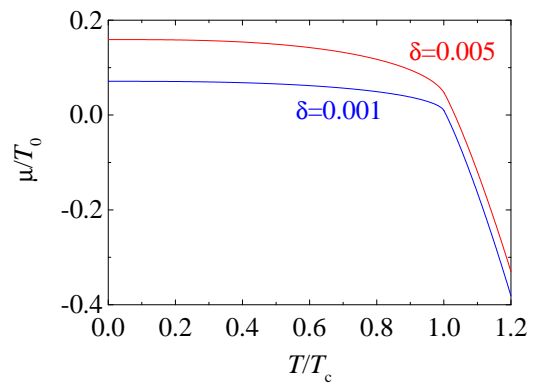}%
\end{center}
\caption{ Chemical potential
as a function of $T/T_c$
for $\delta_{p}\!=\! 0.001$ and $0.005$.
}
\label{fig:7}
\end{figure}
\begin{figure}[t]
\begin{center}
\includegraphics[width=7cm]{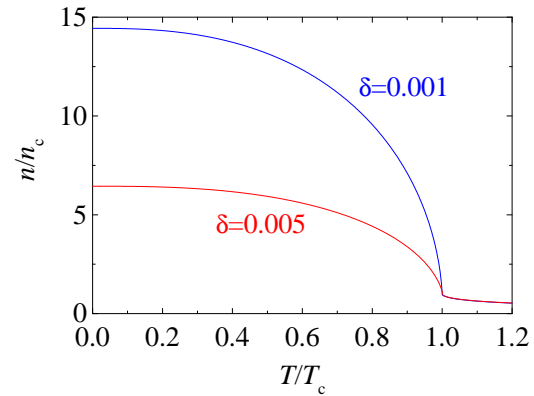}%
\end{center}
\caption{ Normalized particle density $n/n_{\rm c}$ 
as a function of $T/T_c$
for $\delta_{p}\!=\! 0.001$ and $0.005$.
}
\label{fig:8}
\end{figure}
Let us move onto the overall temperature dependences of 
various thermodynamic quantities.
Figure 5 plots specific heat $C_{p}$
over $0.6T_{c}\!\leq\! T\!\leq\! 1.2T_{c}$
for $\delta_{p}\!=\! 0.001$ and $0.005$.
A remarkable enhancement of $C_{p}$
near $T_{c}$ is clearly seen.
Indeed, the curves show a divergent behavior 
$\propto\! |T\!-\! T_{c}|^{-1/2}$ on both sides
near $T_{c}$.
It is completely different from the discontinuous
behavior predicted by the
Landau theory of second-order transition
and caused by the non-analytic nature of the order-parameter
expansion near $T_{c}$.
The sharp heat absorption/emission near $T_{c}$
may be realized in terms of the proximity to the ideal Bose
gas under constant pressure
which suddenly falls into the ground state of
volume zero below $T_{0}$.
The weak repulsive interaction weakens the singularity 
of the ideal Bose gas at $T\!=\! T_{0}$ over a finite range close
to $T_{c}$.
This sharp absorption/emission of heat near $T_{c}$
in the present system is in qualitative agreement
with the observation on $^{4}$He.\cite{Keller69}
Figure 6 shows the normalized condensate density $n_{0}/n$ 
as a function of temperature, where a steep change 
just below $T_{c}$ 
can also be seen clearly.
Although not plotted here,
the normalized superfluid density $\rho_{\rm s}(T)/mn(T)$
is almost indistinguishable in the present scale from $n_{0}(T)/n(T)$
for both $\delta_{p}\!=\! 0.001$ and $0.005$.
Figure 7 displays temperature dependence of the chemical potential $\mu$.
The curves flatten out rapidly below $T_{c}$
reflecting proximity to the ground state.
Finally, Fig.\ 8 plots the particle density $n$ as a function of
temperature for $\delta_{p}\!=\! 0.001$ and $0.005$.
Since the superfluid component contributes less to the
pressure than the normal component in the weak-coupling region,
a constant pressure can only be sustained at 
lower temperatures by increasing the particle density.
Thus, $n(T)$ increases monotonically 
as $T$ is lowered from $T_{c}$.
The leading-order expression of $n( 0)$
is given by $n(0)/n_{\rm c}\!
=\! [\zeta(\frac{5}{2})^{3/5}/\zeta(\frac{5}{2})]\delta_{p}^{-1/2}
\!=\!0.457\delta_{p}^{-1/2}$,
implying a smaller enhancement of $n(0)$ for a stronger interaction.
In this respect, it is worth pointing out that
$n(T)$ of superfluid $^{4}$He displays a maximum at $T_{c}$.\cite{Keller69}
Whether $n(0)/n(T_{c})$ may be reduced beyond $1$ or not
in the strong-coupling region remains a challenging problem 
to be clarified theoretically.

\section{Summary}

We have constructed a mean-field theory 
for Bose-Einstein condensates.
The basic equations to determine
the equilibrium are given by Eqs.\
(\ref{SD})-(\ref{hatH}), (\ref{BdG}), and (\ref{GF}).
They can be applicable to general non-uniform systems
such as the trapped atomic gases with or without vortices.
The present theory has a manifest advantage
over the HFB and Shohno theories
that it simultaneously satisfies the Hugenholtz-Pines theorem
and various conservation laws.
It will be useful in clarifying both 
equilibrium and dynamical properties of condensed
Bose gases over the whole temperature range.
The conserving gapless theory is expected to provide at least a qualitative
description of weakly interacting BEC.
Especially, it will shed a new light on the temperature
dependences of the collective modes in BEC.
It may also form a starting point
for a unified description of BEC over the whole interaction
strengths from the trapped weakly interacting
atomic gases to superfluid $^4$He.

The theory is then applied to a homogeneous weakly interacting Bose gas
with $s$-wave scattering length $a$ under constant density $n$
in Sec.\ III and under constant pressure $p$ in Sec.\ IV.
The order-parameter expansion near $T_{c}$ is found not of the
Landau-Ginzburg type but non-analytic by nature due to the
divergence of the Bose distribution function at zero energy,
as shown explicitly in Sec.\ IIIB.
This non-analyticity makes the superfluid transition first-order
and also brings unique temperature dependences of 
various thermodynamic quantities near $T_{c}$
which are quite different from those of the Landau
theory of second-order transition.
The transition temperature $T_c$ under constant $n$
is found to increase from the
ideal Bose gas value $T_0$ as $T_{c}/T_{0}
\!=\! 1\!+\! 2.33an^{1/3}T_{0}$, whereas 
it decreases under constant pressure as
$T_{c}/T_{0}\!=\! 1\!-\! 3.84a(mp/2\pi\hbar^{2})^{1/5}$
We have also clarified overall temperature dependences of 
basic thermodynamic quantities in Secs.\ IIID and IV.
These predictions are then compared with those 
of the HFB\cite{GA59,Luban62,RS69,Griffin96} 
and Shohno\cite{Shohno64,RS69} theories in Sec.\ \ref{subsec:HFB}.
Those theories also predict a first-order transition
and  an enhancement of $T_{c}$.

We finally comment on the first-order superfluid transition of
the present mean-field theory. 
As mentioned above, it is a feature inherent in the mean-field
theories of BEC stemming from the non-analyticity of the
Bose distribution funtion at zero energy.
However, it is in contradiction to
the observation of superfluid $^{4}$He where the transition is
continuous.
It has been pointed out by Reatto and Straley \cite{RS69} 
that the mean-field transition
would be continuous 
if the change of the single-particle dispersion at $T_{c}$
were from $k^{\epsilon}$ to $k^{{\epsilon}/{2}}$
with $\epsilon\!<\!\frac{3}{2}$ ($k$: wave number).
The behavior $k^{\epsilon}$ with $\epsilon\!<\!\frac{3}{2}$
above $T_{c}$ may only be possible by including correlations and/or 
fluctuations.
Thus, we encounter the question of whether the 
correlations and/or 
fluctuations are essential even in the limit $a\!\rightarrow\! 0$
to transform the nature of the superfluid transition.
If not so, the transition is expected to change its character
from first-order into continuous at an intermediate value of $a$. 
Thus, the predictions of the mean-field theory
raise the questions on: (i)
the relative importance of the single-particle and collective 
excitations in BEC as a function of the interaction strength;
(ii) the role of the single-particle 
excitations in superfluid $^4$He above and below $T_{c}$.
It is worth reminding that the transition of 
the ideal Bose gas under constant
pressure is strongly first-order.
Hence it may not be so unreasonable to expect
that the first-order transition persists into the region
of a finite interaction.

Thus far, microscopic theoretical studies
on condensed Bose systems seem to
have been carried out separately in two 
limiting cases with completely different approaches: 
(a) the weak-coupling regime where
certain model Hamiltonians have been investigated
based on the field theory;\cite{Andersen04,Yukalov04}
(b) the strong coupling regime where interests
have been focused on describing
superfluid $^{4}$He quantitatively
based on some variational wave functions,\cite{Feynman54,
McMillan65,Feenberg69}
lattice-gas models which can be mapped onto
spin systems,\cite{MM56}
and quantum Monte-Carlo calculations.\cite{Ceperley95}
We hence need systematic investigations based 
on a model Hamiltonian from weak- to strong-coupling regimes,
which may provide us more profound understanding 
on condensed Bose systems and shed a new light on
the nature of the still mysterious lambda transition 
of superfluid $^{4}$He.\cite{Lipa96}

\acknowledgements
The author is grateful to V.\ I.\ Yukalov and
Z.\ Te\v sanovi\'c for stimulating discussions.
This work is supported by
the 21st century COE program ``Topological Science and Technology,'' 
Hokkaido University.

\appendix

\section{Feynman rules in Nambu space}

The bare perturbation expansion
of the condensed Bose systems is usually carried out
with respect to the condensate wave function $\Psi({\bf r})$ and
the non-interacting Green's function:
\begin{equation}
G^{(0)}({\bf r}_{1}\tau_{1},{\bf r}_{2}\tau_{2})
\!\equiv\! -\langle
T_{\tau}\phi({\bf r}_{1}\tau_{1})\phi^{\dagger}({\bf r}_{2}\tau_{2})
\rangle_{0}  \, .
\end{equation}
The anomalous (i.e., off-diagonal) Green's function is not necessary
before the renormalization,
and we have well-established Feynman rules for this purpose.
We will show that the same rules can be used with a slight modification
to the perturbation expansion in terms of the Nambu matrix of Eq.\
(\ref{hatG}).
This extension will be performed in exactly the same way
as the extension from the normal to the superconducting 
Fermi systems.\cite{Kita96}

Let us specifically consider the following term
in the interaction:
\begin{equation}
H^{(3)}\equiv U_{11'}\!\left(
\Psi^{*}_{1}\phi^{\dagger}_{1'}
\phi_{1'}\phi_{1}
+\phi^{\dagger}_{1}\phi^{\dagger}_{1'}
\phi_{1'}\Psi_{1}\right)  ,
\label{H_3a}
\end{equation}
where ${\cal U}_{11'}\!\equiv\!{\cal U}({\bf r}_{1}\!-\!{\bf r}_{1}')$,
$\phi_{1}\!\equiv\!\phi({\bf r}_{1})$, etc., 
and summations over the repeated indices are implied. 
Using the normal-ordering operator $N$, we first rewrite 
Eq.\ (\ref{H_3a}) as
\begin{subequations}
\label{H_3}
\begin{equation}
H^{(3)}= {\cal U}_{11'}N\!\left(
\Psi^{*}_{1}\phi_{1}\phi^{\dagger}_{1'}
\phi_{1'}
+\phi^{\dagger}_{1}\Psi_{1}
\phi^{\dagger}_{1'}
\phi_{1'}\right)  .
\label{H_3b}
\end{equation}
With the operator $N$ in front, Eq.\ (\ref{H_3a}) is 
further transformed into
\begin{equation}
H^{(3)}= {\cal U}_{11'}N\!\left(
\frac{1}{2}
\vec{\Psi}^{\dagger}_{1}\vec{\phi}_{1}
\frac{1}{2}
\vec{\phi}^{\,\dagger}_{1'}
\vec{\phi}_{1'}
+\frac{1}{2}
\vec{\phi}^{\,\dagger}_{1}\vec{\Psi}_{1}
\frac{1}{2}
\vec{\phi}^{\,\dagger}_{1'}
\vec{\phi}_{1'}\right)  ,
\label{H_3c}
\end{equation}
\end{subequations}
where $\vec{\phi}$ and $\vec{\Psi}$ are the spinors of Eq.\ (\ref{spinors}).
Comparing Eqs.\ (\ref{H_3b}) and (\ref{H_3c}),
we observe the following correspondence:
\begin{equation}
\phi_{1}^{\dagger} \leftrightarrow 
\vec{\phi}_{1}^{\,\dagger} \, ,
\hspace{3mm}
\Psi_{1}^{*}\leftrightarrow \vec{\Psi}_{1}^{\dagger} \, ,
\hspace{3mm}
\phi_{1}^{\dagger}\phi_{1} \leftrightarrow 
\frac{1}{2}\vec{\phi}_{1}^{\,\dagger}\vec{\phi}_{1} \, ,
\hspace{2mm}
{\rm etc.}
\label{CorrespondNS}
\end{equation}
Let us perform a perturbation expansion in the imaginary time
domain with respect to (\ref{H_3c}).
We hence include $\tau_{i}$ in the index $i$, i.e., 
$i\!\equiv\!{\bf r}_{i}\tau_{i}$
and $i'\!\equiv\!{\bf r}_{i}'\tau_{i}$.
Consider an $n$th-order term in the expansion and
take a contraction of $\vec{\phi}_{i}^{\,\dagger}$
with $\frac{1}{2}\vec{\phi}_{j}^{\,\dagger}\vec{\phi}_{j}$.
We then find that 
$\vec{\phi}_{j}^{\,\dagger}$ and $\vec{\phi}_{j}$ 
in $\frac{1}{2}\vec{\phi}_{j}^{\,\dagger}\vec{\phi}_{j}$
contribute equivalently to the contraction, as shown easily
by writing it down
with respect to $\phi_{i}$,
$\phi_{i}^{\dagger}$, $\phi_{j}$,
and $\phi_{j}^{\dagger}$;
the statement is valid even if we formally retain the anomalous averages.
Because of this fact, we can introduce the rule:
Contract $\vec{\phi}_{i}^{\,\dagger}$ only
with $\vec{\phi}_{j}$
in $\frac{1}{2}\vec{\phi}_{j}^{\,\dagger}\vec{\phi}_{j}$
and multiply the contribution by $2$.
This yields $\vec{\phi}_{j}^{\,\dagger}
\langle T_{\tau} N\vec{\phi}_{j}
\vec{\phi}_{i}^{\,\dagger}\rangle_{0}$, 
where $N$ is effective only when 
$i$ and $j$ belong to the same interaction line
according to the definition in Eq.\ (\ref{H_3}).
Note $\langle T_{\tau} \vec{\phi}_{j}
\vec{\phi}_{i}^{\,\dagger}\rangle_{0}\!=\!
-\hat{\tau}_{3}\hat{G}_{ji}^{(0)}$ as realized from Eq.\ (\ref{hatG}).
The rule allows us to perform a perturbation expansion in
Nambu space by using only the Feynman diagrams for $G^{(0)}_{ij}$.
The following two points should be supplemented:
(a) A factor $\frac{1}{2}$ remains uncancelled
for each closed particle line,
due to the fact that we have to fix
the starting pair $\frac{1}{2}\vec{\phi}_{i}^{\,\dagger}
\vec{\phi}_{i}$
for each closed line.
(b) The contraction of $\vec{\phi}_{i}$ with the last 
$\vec{\phi}_{k}^{\,\dagger}$ in the closed particle line
may be transformed as
\begin{equation}
\frac{1}{2}
\langle T_{\tau} N \vec{\phi}_{k}^{\,\dagger}\hat{A}_{ki}
\vec{\phi}_{i}\rangle_{0}
=\frac{1}{2}{\rm Tr}
\hat{A}_{ki}\langle T_{\tau} N\vec{\phi}_{i}
\vec{\phi}_{k}^{\,\dagger}\rangle_{0}\, ,
\end{equation}
where $\hat{A}_{ki}$ is a Nambu matrix composed of the contractions
connecting $k$ with $i$.
(c) As for the open line starting from 
$\frac{1}{2}\vec{\phi}_{i}^{\,\dagger}
\vec{\Psi}_{i}$ and ending at
$\frac{1}{2}\vec{\Psi}_{k}^{\,\dagger}
\vec{\phi}_{k}$, 
the two factors $\frac{1}{2}$ remain uncancelled.
These rules are valid even after including the interactions
other than Eq.\ (\ref{H_3}).

In summary, we can carry out a perturbation expansion in Nambu space 
by modifying the Feynman rules for $G^{(0)}$, $\Psi$, and $\Psi^{*}$
as follows:
(i) $G^{(0)}_{ij}\!\rightarrow\!\hat{\tau}_{3}\hat{G}^{(0)}_{ij}$,
$\Psi_{i}\!\rightarrow\!\frac{1}{2}\vec{\Psi}_{i}$,
and $\Psi_{i}^{*}\!\rightarrow\!\frac{1}{2}\vec{\Psi}_{i}^{\dagger}$;
(ii) $\frac{1}{2}{\rm Tr}$ for every closed particle line.
Note that the operator $N$ in
$\langle T_{\tau} N\vec{\phi}_{i}
\vec{\phi}_{k}^{\,\dagger}\rangle_{0}$
transforms
into the matrix of Eq.\ (\ref{1_n})
in the frequency space. 

Let us write down the 
contributions of the diagrams in Fig.\ \ref{fig:1}
based on the above rules 
with $\hat{G}^{(0)}\!\rightarrow\!\hat{G}$. 
This yields $\Phi_{\rm HFB}$.
We then modify it
slightly as described below Eq.\ (\ref{Phi}).
We thereby obtain $\Phi$ of Eq.\ (\ref{Phi}).

\section{Conservation laws}

It is shown that the $\Phi$ derivative approximation
defined by Eqs.\ (\ref{SigmaDef}) and (\ref{eta}) automatically
satisfies various conservation laws.

\subsection{Definitions}

Since we work in the real-time domain, 
we need modifications of various definitions.
The Green's function is now defined by
\begin{equation}
\hat{G}(1,2)\equiv
-
\frac{i}{\hbar}\hat{\tau}_{3} \langle T
\vec{\phi}(1)
\vec{\phi}^{\,\dagger}(2)\rangle
\, ,
\label{hatG(t)}
\end{equation}
with $1\!\equiv\! {\bf r}_{1}t_{1}$. It obeys the equation of motion:
\begin{subequations}
\label{EM}
\begin{equation}
i\hbar \frac{\partial\hat{G}(1,1')}{\partial t_{1}}
\!-\!\hat{\tau}_{3}K_{1}\hat{G}(1,1')
\!=\!\hat{1}\delta(1,1')\!+\!\hat{\Sigma}(1,\bar{2})\hat{G}(\bar{2},1')
\, ,
\label{Dyson(t)-l}
\end{equation}
or equivalently,
\begin{equation}
-i\hbar \frac{\partial \hat{G}(1,1')}{\partial t_{1'}}
\!-\!K_{1'}\hat{G}(1,1')\hat{\tau}_{3}
\!=\!\hat{1}\delta(1,1')\!+\!\hat{G}(1,\bar{2})\hat{\Sigma}(\bar{2},1')
\, .
\label{Dyson(t)-r}
\end{equation}
Here $K_{1}$ is given by Eq.\ (\ref{H^(0)}), 
and integration
over the barred index $\bar{2}$ is implied.
Next, the condensate wave function satisfies
\begin{equation}
i\hbar \frac{\partial}{\partial t_{1}}\vec{\Psi}(1)
-K_{1}\hat{\tau}_{3}\vec{\Psi}(1)
=\hat{\tau}_{3}\vec{\eta}(1) \, ,
\label{GP(t)}
\end{equation}
or equivalently,
\begin{equation}
-i\hbar \frac{\partial}{\partial t_{1}}\vec{\Psi}^{\dagger}(1)
-K_{1}\vec{\Psi}^{\dagger}(1)\hat{\tau}_{3}
=\vec{\eta}^{\,\dagger}(1)\hat{\tau}_{3} \, .
\label{GP(t)-r}
\end{equation}
\end{subequations}
Here $\vec{\Psi}$ is given in Eq.\ (\ref{Psi}),
and $\vec{\eta}(1)$ denotes
\begin{equation}
\vec{\eta}(1)\equiv \left[
\!
\begin{array}{c}
\vspace{1mm}
\eta(1)
\\
\eta^{*}(1)
\end{array}
\right]
\, .
\label{eta-vec}
\end{equation}
The key quantities in Eq.\ (\ref{EM}) are 
$\hat{\Sigma}$ and $\eta$.
In the $\Phi$-derivative approximation\cite{HM65} adopted here,
they are determined from a functional 
$\Phi\!=\!\Phi(\hat{G},{\Psi},{\Psi}^{*})$ by
\begin{subequations}
\label{Sigma-eta-def}
\begin{equation}
\hat{\Sigma}(1,1')
=-2i\frac{\delta \Phi}{\delta \hat{G}(1',1)}\, ,
\label{Sigma-def}
\end{equation}
\begin{equation}
\eta(1)=\hbar\frac{\delta \Phi}{\delta \Psi^{*}(1)}
\, .
\label{eta-def}
\end{equation}
\end{subequations}
Note that this $\Phi$ is different in definition from those
of Baym\cite{Baym62} and Hohenberg and Martin\cite{HM65}
by factor $i$, i.e., $\Phi\!=\!i\Phi_{\rm Baym}\!=\!i\Phi_{\rm HM}$.
The present definition has an advantage that $\Phi$
becomes real within the mean-field approximation.
A comment is also necessary on Eq.\ (\ref{hatG(t)}).
When $t_{2}\!=\!t_{1}$,
the operator $\phi^{\dagger}$ 
in $\hat{G}(1,2)$ should be placed
to the left of $\phi$ by definition.
This procedure is
in accordance with the treatment on the equal-time
operators in the calculation of $\Phi$.
Put it another way, 
the equal time implies $t_{2}\!=\!t_{1+}$
for the $(11)$ component of $\hat{G}$ and $t_{2}\!=\!t_{1-}$
for the $(22)$ component. Thus,
$t_{2}$ is not exactly equal to $t_{1}$ for the diagonal elements
of $\hat{G}$.

\subsection{Identites}

With Eq.\ (\ref{Sigma-eta-def}), we can prove several identities
by exactly 
following the argument of Baym.\cite{Baym62}

Suppose we perform the gauge transformation:
\begin{equation}
\left\{\!\!
\begin{array}{l}
\vspace{1mm}
\hat{G}(1,2)\rightarrow {\rm e}^{i\chi(1)\hat{\tau}_{3}}
\hat{G}(1,2) {\rm e}^{-i\chi(2)\hat{\tau}_{3}} 
\\
{\Psi}(1)\rightarrow {\rm e}^{i\chi(1)}
{\Psi}(1)
\end{array} \right. \!\! ,
\label{gauge}
\end{equation}
where $\chi(1)$ is an arbitrary real function.
The corresponding variations 
in $\hat{G}$ and ${\Psi}$ are
given to the leading order by
\begin{equation}
\left\{\!\!
\begin{array}{l}
\vspace{1mm}
\delta\hat{G}(1,2)= i[\chi(1)\hat{\tau}_{3}\hat{G}(1,2)
-\hat{G}(1,2)\chi(2)\hat{\tau}_{3}]
\\
\delta{\Psi}(1)= i
\chi(1){\Psi}(1)
\end{array} \right. \!\! ,
\label{dG-gauge}
\end{equation}
respectively. However, $\Phi$ is clearly invariant under
Eq.\ (\ref{gauge}), i.e., $\delta\Phi\!=\!0$. This yields
\begin{equation}
\frac{i\hbar}{2}{\rm Tr}
\Sigma(\bar{1},\bar{2})\delta\hat{G}(\bar{2},\bar{1})
+
\delta{\Psi}^{*}(\bar{1})\eta(\bar{1})
+\delta{\Psi}(\bar{1})\eta^{*}(\bar{1}) =0\, .
\label{dPhi}
\end{equation}
Substituting Eq.\ (\ref{dG-gauge}) into Eq.\ (\ref{dPhi})
and recalling $\chi(1)$ is arbitrary, we obtain
\begin{equation}
\frac{1}{2}{\rm Tr}\hat{\tau}_{3}\!\bigl[\hat{\Sigma}(1,\bar{2})
\hat{G}(\bar{2},1)-\hat{G}(1,\bar{2})
\hat{\Sigma}(\bar{2},1)\hspace{-0.3mm}\bigr]
\!-\frac{i}{\hbar}\vec{\Psi}^{\dagger}\hspace{-0.4mm}(1)  \hat{\tau}_{3}
\vec{\eta}(1) \!=\! 0 \hspace{0.2mm} ,
\label{gauge-identity}
\end{equation}
with $\vec{\Psi}^{\dagger}\!\equiv\!(
\Psi^{*}\, \Psi)$.

We next consider the Galilean transformation:
\begin{equation}
\left\{\!\!
\begin{array}{l}
\vspace{1mm}
\hat{G}(1,2)\rightarrow 
\hat{G}({\bf r}_{1}\!+\!{\bf R}(t_{1}),t_{1};
{\bf r}_{2}\!+\!{\bf R}(t_{2}),t_{2})  
\\
{\Psi}(1)\rightarrow 
{\Psi}({\bf r}_{1}\!+\!{\bf R}(t_{1}),t_{1})
\end{array} \right. \!\!\! .
\label{Galilean}
\end{equation}
The corresponding leading-order variations in $\hat{G}$
and $\Psi$ are given by
\begin{equation}
\left\{\!\!
\begin{array}{l}
\vspace{1mm}
\delta\hat{G}(1,2)=
[{\bf R}(t_{1})\!\cdot\!{\bf\nabla}_{\! 1}
+{\bf R}(t_{2})\!\cdot\!{\bf\nabla}_{2}]
\hat{G}(1,2)
\\
\delta{\Psi}(1)= {\bf R}(t_{1})\!\cdot\!{\bf\nabla}_{\! 1}{\Psi}(1)
\end{array} \right. \!\!\! .
\label{dG-Galilean}
\end{equation}
However, $\Phi$ remains invariant since
it is a quantity obtained by integrations
over all the space-time variables. 
We hence conclude that Eq.\ (\ref{dPhi}) 
holds also in this case.
Substituting Eq.\ (\ref{dG-Galilean}) into Eq.\ (\ref{dPhi})
and recalling ${\bf R}(t)$ is arbitrary,
we obtain
\begin{equation}
\!\int\!{\bf S}(1)\, d{\bf r}_{1}={\bf 0} \, ,
\label{Galilean-identity}
\end{equation}
with
\begin{eqnarray}
&&\hspace{-9mm}
{\bf S}(1)\equiv
\frac{1}{4}({\bf\nabla}_{\! 1}\!-\!{\bf\nabla}_{\! 1'})
\bigl\{ {i\hbar}\,{\rm Tr}\bigl[
\hat{G}(1,\bar{2}) \hat{\Sigma}(\bar{2},1')- 
\hat{\Sigma}(1,\bar{2})
\nonumber \\
&&\hspace{2mm}
\times
\hat{G}(\bar{2},1')\bigr]-2\vec{\Psi}^{\dagger}(1')\vec{\eta}(1)
\bigr\}\bigr|_{1'=1} \, .
\label{vecS}
\end{eqnarray}
Here terms with ${\bf\nabla}_{\! 1}\hat{\Sigma}(1,2)$
and ${\bf\nabla}_{\! 1}\vec{\eta}(1)$ have been obtained 
through integrations
by parts.

We finally consider the ``rubber-clock'' transformation
of Baym:\cite{Baym62}
$t\!\rightarrow\!\theta(t)\!\equiv\!t\!+\! f(t)$ with 
$f(\pm\infty)\!=\! 0$.
It yields
\begin{equation}
\left\{\!\!
\begin{array}{l}
\vspace{1mm}
\hat{G}(1,2)\rightarrow 
\displaystyle
\left(\!\frac{d\theta_{1}}{dt_{1}}\!\right)^{\!\!1/4}
\hat{G}({\bf r}_{1}\theta_{1},
{\bf r}_{2}\theta_{2})  
\left(\!\frac{d\theta_{2}}{dt_{2}}\!\right)^{\!\!1/4}
\\
\displaystyle
{\Psi}(1)\rightarrow 
\left(\!\frac{d\theta_{1}}{dt_{1}}\!\right)^{\!\!1/4}
{\Psi}({\bf r}_{1}\theta_{1})
\end{array} \right.\!\!\! .
\label{t-tau}
\end{equation}
The factor $(d\theta/dt)^{1/4}$ 
compensates the Jacobian for
$t\!\rightarrow\!\theta$, thereby making 
the integration of $H_{\rm int}(t)$ over $t$ invariant
in form through the change of variables.
The corresponding leading-order variations in $\hat{G}$
and $\Psi$ are given by
\begin{equation}
\left\{\!\!
\begin{array}{l}
\vspace{1mm}
\displaystyle
\delta\hat{G}(1,2)=\!\left(\frac{f'_{1}\!+\!f'_{2}}{4}
\!+\! f_{1}\frac{\partial}{\partial t_{1}}
\!+\! f_{2}\frac{\partial}{\partial t_{2}}\right)\!
\hat{G}(1,2)
\\
\displaystyle
\delta{\Psi}(1)= 
\!\left(\frac{f'_{1}}{4}
\!+\! f_{1}\frac{\partial}{\partial t_{1}}\!\right)\!{\Psi}(1)
\end{array} \right. \!\!\! ,
\label{dG-time}
\end{equation}
with $f_{j}\!\equiv\!f(t_{j})$.
However, $\Phi$ remains invariant. Hence Eq.\ (\ref{dPhi}) holds
also in this case.
Substituting Eq.\ (\ref{dG-time}) into Eq.\ (\ref{dPhi})
and recalling $f(t)$ is arbitrary,
we obtain the third identity:
\begin{eqnarray}
&&\hspace{-5mm}
\frac{d\langle H_{\rm int}\rangle_{1}}{d t_{1}}
=\frac{1}{2}\int\!d{\bf r}_{1}\frac{\partial}{\partial t_{1}}
\bigl\{i\hbar{\rm Tr}\bigl[\hat{G}(1,\bar{2})
\hat{\Sigma}(\bar{2},1')\!+\! \hat{\Sigma}(1',\bar{2})
\nonumber \\
&&\hspace{2mm}
\times
\hat{G}(\bar{2},1)\bigr]\!+\!\vec{\Psi}^{\dagger}(1)\vec{\eta}(1')\!+\!
\vec{\eta}^{\,\dagger}(1')\vec{\Psi}(1) \bigr\}_{1'=1}  \, ,
\label{time-identity}
\end{eqnarray}
where $\langle H_{\rm int}\rangle_{1}$ is {\em defined by}
\begin{eqnarray}
&&\hspace{-6mm}
\langle H_{\rm int}\rangle_{1}=
\frac{1}{8}\int\!d{\bf r}_{1}
\bigl\{i\hbar\,{\rm Tr}
\bigl[\hat{G}(1,\bar{2})
\hat{\Sigma}(\bar{2},1)\!+\! \hat{\Sigma}(1,\bar{2})\hat{G}(\bar{2},1)\bigr]
\nonumber \\
&&\hspace{12mm}
+\vec{\Psi}^{\dagger}(1)\vec{\eta}(1)
\!+\!\vec{\eta}^{\,\dagger}(1)\vec{\Psi}(1)
 \bigr\}\, .
\label{Hint}
\end{eqnarray}
Equation (\ref{Hint}) indeed corresponds to the interaction 
energy of the system.
This can be checked as follows:
(i) write down the equations of motion for
$\phi$ and $\Psi$ in terms of the field operators
and the condensate wave function; (ii) construct the interaction
energy by adding those equations appropriately; and (iii) 
transform the one-particle energy in the equation
by using Eq.\ (\ref{EM}).

Equations (\ref{gauge-identity}), (\ref{Galilean-identity}),
and (\ref{time-identity}) are the basic identities to be used below.

\subsection{Conservation laws}

Let us operate $\frac{1}{2}{\rm Tr}\hat{\tau}_{3}$,
$-\frac{1}{2}{\rm Tr}\hat{\tau}_{3}$, 
and $-\frac{i}{\hbar}\vec{\Psi}^{\dagger}(1)$ 
to Eqs.\ (\ref{Dyson(t)-l}),
(\ref{Dyson(t)-r}), and (\ref{GP(t)}),
respectively.
Adding the resulting three equations, setting $1'\!=\! 1$, and noting
Eq.\ (\ref{gauge-identity}), we obtain the particle conservation law as
\begin{equation}
\frac{\partial n(1)}{\partial t_{1}}+{\bf\nabla}_{\! 1} {\bf j}(1)=0 \, ,
\label{pc}
\end{equation}
where $n(1)$ and ${\bf j}(1)$ are defined by
\begin{equation}
n(1)=\rho(1,1) \, ,
\label{number}
\end{equation}
\begin{equation}
{\bf j}(1)=\frac{-i\hbar}{2m}({\bf\nabla}_{\! 1}\!-\!
{\bf\nabla}_{\! 1'})\rho(1,1')\bigr|_{1'=1} \, ,
\label{current}
\end{equation}
with 
\begin{equation}
\rho(1,1')\equiv \langle \phi^{\dagger}(1')\phi(1)\rangle
+{\Psi}(1){\Psi}^{*}(1') \, .
\label{rho-App}
\end{equation}
This $\rho(1,1')$ is the one-body density matrix
appearing in the evaluation of every one-particle operator.

To obtain the momentum conservation law,
we operate $-\frac{i\hbar}{4m}{\rm Tr}
({\bf\nabla}_{\! 1}\!-\!{\bf\nabla}_{\! 1'})$,
$\frac{i\hbar}{4m}{\rm Tr}
({\bf\nabla}_{\! 1}\!-\!{\bf\nabla}_{\! 1'})$, 
and $-\frac{1}{2m}\vec{\Psi}^{\dagger}(1)\hat{\tau}_{3}{\bf\nabla}_{\! 1}$
to Eqs.\ (\ref{Dyson(t)-l}),
(\ref{Dyson(t)-r}), and (\ref{GP(t)}), respectively.
We also multiply Eq.\ (\ref{GP(t)-r}) from the right by
$\frac{1}{2m}\hat{\tau}_{3}{\bf\nabla}_{\! 1}\vec{\Psi}(1)$.
Adding the four equations and setting $1'\!=\! 1$, we obtain
\begin{equation}
\frac{\partial}{\partial t_{1}}{\bf j}(1)
+\frac{n(1)}{m}{\bf\nabla}_{\! 1}V(1)
+\frac{1}{m}{\bf\nabla}_{\! 1}\cdot{\bf \Pi}(1)
=\frac{1}{m}{\bf S}(1) \, .
\label{momentum-eq}
\end{equation}
Here ${\bf j}(1)$, $n(1)$, and ${\bf S}(1)$ are given by 
Eqs.\ (\ref{current}), (\ref{number}), and (\ref{vecS}), respectively,
and ${\bf \Pi}(1)$ is a tensor defined by
\begin{equation}
\Pi_{\alpha\beta}(1)=-\frac{\hbar^{2}}{4m}
(\nabla_{\! 1\alpha}\!-\!\nabla_{\! 1'\alpha})
(\nabla_{\! 1\beta}\!-\!\nabla_{\! 1'\beta})\rho(1,1')\bigr|_{1'=1} \, .
\end{equation}
Integrating Eq.\ (\ref{momentum-eq}) over ${\bf r}_{1}$ and
noting Eq.\ (\ref{Galilean-identity}), we obtain
the conservation law for the total momentum as
\begin{equation}
\frac{\partial}{\partial t_{1}}\int\!{\bf j}(1)\,d{\bf r}_{1}
=-\frac{1}{m}\int\! n(1){\bf\nabla}_{\! 1}V(1)\,d{\bf r}_{1}
\, .
\label{Momentum-conservation}
\end{equation}

We finally consider the energy conservation law. 
The time derivative of the interaction energy is given by
Eq.\ (\ref{time-identity}). 
The terms in the curly bracket of Eq.\ (\ref{time-identity})
can be transformed into an expression
of using the left-hand sides of Eq.\ (\ref{EM}).
We then find that 
the terms with the operator $i\hbar\frac{\partial}{\partial t_{1'}}$
cancels out upon applying $\frac{\partial}{\partial t_{1}}$
in front of the curly bracket.
We thereby obtain 
\begin{equation}
\frac{d\langle H_{\rm int}\rangle_{1}}{dt_{1}}
=-\int\!d{\bf r}_{1}K_{1}\frac{\partial}{\partial t_{1}}
\rho({\bf r}_{1}t_{1},{\bf r}_{1}'t_{1})|_{{\bf r}_{1}'={\bf r}_{1}}
\, ,
\label{dHint-dt}
\end{equation}
with $\rho(1,1')$ defined by Eq.\ (\ref{rho-App}).
When $K$ has no time dependence as given by
Eq.\ (\ref{H^(0)}),
Eq.\ (\ref{dHint-dt}) directly leads to 
$\langle K\!+\!H_{\rm int}\rangle\!=\! {\rm constant}$.
Following Kadanoff and Baym,\cite{KB62} we may alternatively
consider $V$ as a small time-dependent external field and 
trace the time variation of the 
internal energy $\langle K^{(0)}\!+\! H_{\rm int}\rangle$,
where $K^{(0)}\!\equiv\!-\frac{{\hbar}^{2}}{2m}\nabla^{2}$.
Using Eq.\ (\ref{pc}) and performing integration by parts
with respect to ${\bf r}_{1}$, we obtain
\begin{equation}
\frac{d\langle K^{(0)}\!+\! H_{\rm int}\rangle_{1}}{dt_{1}}
=-\int {\bf j}(1)\cdot{\bf\nabla}_{\! 1} V(1) \,d{\bf r}_{1} \, .
\label{energy-conservation}
\end{equation}

Thus, we have seen that
the conservation laws are automatically
obeyed as Eq.\ (\ref{pc}), (\ref{Momentum-conservation}), 
and (\ref{energy-conservation})
in the $\Phi$ derivative approximation of
Eq.\ (\ref{Sigma-eta-def}).

\section{The IRK theory}

It will be shown that 
the thermodynamic functional
presented by Ivanov, Riek, and Knoll\cite{IRK05} (IRK)
leads to exactly the 
same thermodynamic properties
for the weakly interacting Bose gases
as the present theory.

The IRK self-energy $\hat{\Sigma}^{({\rm IRK})}$ is calculated from 
Eq.\ (\ref{Phi-IRK}) by Eq.\ (\ref{SigmaDef}).
The result can be written as 
\begin{equation}
\hat{\Sigma}^{({\rm IRK})}({\bf r},{\bf r}')
=
\left[
\begin{array}{cc}
\vspace{1mm}
\Sigma({\bf r},{\bf r}') &-\Delta^{\!({\rm IRK})}({\bf r},{\bf r}')
\\
\Delta^{\!({\rm IRK})*}({\bf r},{\bf r}') &-\Sigma^{*}({\bf r},{\bf r}')
\end{array}
\right] ,
\label{hatSigma-IRK}
\end{equation}
where $\Sigma$ is given by Eq.\ (\ref{Sigma}),
and $\Delta^{\!({\rm IRK})}$ denotes
\begin{equation}
\Delta^{\!({\rm IRK})}({\bf r},{\bf r}')
 = {\cal U}({\bf r}\!-\!{\bf r}')
 \tilde{\rho}^{({\rm IRK})}({\bf r},{\bf r}') \, ,
\label{Delta-IRK}
\end{equation}
with
\begin{equation}
\tilde{\rho}^{({\rm IRK})}({\bf r},{\bf r}')
 = \Psi({\bf r})\Psi({\bf r}')-
\langle \phi({\bf r})
\phi({\bf r}')\rangle^{\!({\rm IRK})} \, .
\label{rho2-IRK}
\end{equation}
The minus sign in front of 
$\Delta^{\!({\rm IRK})}$ in Eq.\ (\ref{hatSigma-IRK})
has been introduced to make the coefficient 
$\Delta^{\!({\rm IRK})}_{\bf k}$ for the homogeneous system
positive at low temperatures, in accordance with 
the definition of $\Delta$ in Eq.\ (\ref{Delta}).
Next, the equation for $\Psi({\bf r})$ is obtained from
the thermodynamic functional $\Omega^{\rm IRK}$ by 
$\partial \Omega^{\rm IRK}/\partial \Psi({\bf r})\!=\! 0$.
It is given by Eq.\ (\ref{GP}) with the replacement:
$\Delta\!\rightarrow\!\Delta^{\!({\rm IRK})}$.

In a way parallel to Eqs.\ (\ref{hatH}) and (\ref{BdG}), let us 
introduce the Hamiltonian:
\begin{equation}
\hat{H}^{({\rm IRK})}({\bf r},{\bf r}')\!\equiv\!
\hat{\tau}_{3}K\delta({\bf r}\!-\!{\bf r}')
\!+\!\hat{\Sigma}^{({\rm IRK})}({\bf r},{\bf r}') \, ,
\label{hatH-IRK}
\end{equation}
and construct the eigenvalue problem as
\begin{equation}
\int \hat{H}^{({\rm IRK})}({\bf r},{\bf r}')
\hat{u}_{\nu}^{({\rm IRK})}({\bf r}')
d{\bf r}'
= \hat{u}_{\nu}^{({\rm IRK})}({\bf r}) 
E_{\nu} \hat{\tau}_{3} \, ,
\label{BdG-IRK}
\end{equation}
where $\hat{u}_{\nu}^{({\rm IRK})}$ is defined by
\begin{equation}
\hat{u}_{\nu}^{({\rm IRK})}({\bf r})=
\left[
\begin{array}{cc}
\vspace{1mm}
u_{\nu}({\bf r}) & v_{\nu}({\bf r})
\\
v_{\nu}^{*}({\bf r}) & u_{\nu}^{*}({\bf r})
\end{array}
\right] \, .
\label{uHat-IRK}
\end{equation}
The sign of $v_{\nu}^{*}({\bf r})$ is chosen so that
the coefficient
$v_{\bf k}$ of the homogeneous system is positive,
in accordance with the definition of $v_{\nu}^{*}({\bf r})$
in Eq.\ (\ref{uHat}).
The corresponding Green's function $\hat{G}^{\rm IRK}$ 
is obtained from Eq.\ (\ref{hatG-2}) with the replacement $
\hat{u}_{\nu}\!\rightarrow\!\hat{u}_{\nu}^{({\rm IRK})}$.
Noting Eq.\ (\ref{F})
and the difference between Eqs.\ (\ref{uHat}) and (\ref{uHat-IRK}), 
we now conclude that $\langle \phi({\bf r})
\phi({\bf r}')\rangle^{\!({\rm IRK})}$
in Eq.\ (\ref{rho2-IRK}) satisfies 
$-\langle \phi({\bf r})
\phi({\bf r}')\rangle^{\!({\rm IRK})}\!=\!
\langle \phi({\bf r})
\phi({\bf r}')\rangle$, which implies
$\Delta^{\!({\rm IRK})}\!=\!\Delta$.
Thus, the equations to determine the equilibrium are
exactly the same between the present theory and the IRK theory.
Finally, one can show that the equilibrium thermodynamic potential
of the IRK theory is also given by Eq.\ (\ref{Omega-eq}).
Thus, the two functionals (\ref{Phi}) and
(\ref{Phi-IRK}) yield exactly the same thermodynamic properties
on BEC.

\section{Non-Hermitian eigenvalue problem}

In this Appendix we study possible origins
of the non-Hermitian eigenvalue 
problem in BEC.
It is shown that we have to treat the non-Hermitian
matrix of Eq.\ (\ref{BdG}) necessarily if we impose the two requirements:
(i) the quasiparticle field
obey the Bose statistics; (ii) the quasiparticle eigenstates
be obtained by diagonalizing a matrix.

Let us introduce the quasiparticle field $\gamma_{\nu}$
through
\begin{equation}
\left[
\begin{array}{c}
\vspace{1mm}
\phi_{{\bf r}} \\
-\phi^{\dagger}_{{\bf r}}
\end{array}
\right] 
=
\sum_{\nu} 
\left[
\begin{array}{cc}
\vspace{1mm}
u_{{\bf r}\nu} & -v_{{\bf r}\nu} 
\\
-v_{{\bf r}\nu}^{*} & u_{{\bf r}\nu}^{*}
\end{array}
\right] 
\left[
\begin{array}{c}
\vspace{1mm}
\gamma_{\nu} 
\\
-\gamma_{\nu}^{\dagger}
\end{array}
\right] ,
\label{BT}
\end{equation}
where $\nu$ distinguishes quasiparticle eigenstates, and 
$\phi_{{\bf r}}\!=\!\phi({\bf r})$,
$u_{{\bf r}\nu}\!=\!u_{\nu}({\bf r})$, etc.
Equation (\ref{BT}) may be written symbolically 
in a compact form as
\begin{equation}
\vec{\phi}=\hat{u}\vec{\gamma} \, ,
\end{equation}
where
\begin{equation}
\vec{\phi}\equiv
\left[\!\!
\begin{array}{c}
\phi_{{\bf r}} \\
\phi_{{\bf r}'} \\
\vdots \\
\vspace{1mm}
-\phi_{{\bf r}}^{\dagger} \\
\vspace{1mm}
-\phi_{{\bf r}'}^{\dagger} \\
\vdots
\end{array}
\!\!\right], \hspace{2mm}
\hat{u}\equiv \left[\!\!
\begin{array}{cc}
\vspace{1mm}
\underline{u} \!\! & \!\! -\underline{v}
\\
-\underline{v}^{*} \!\! &\!\! \underline{u}^{*}
\end{array}
\!\! \right], \hspace{2mm} 
\vec{\gamma}\equiv
\left[\!\!
\begin{array}{c}
\gamma_{\nu} \\
\gamma_{\nu'} \\
\vdots \\
\vspace{1mm}
-\gamma_{\nu}^{\dagger} \\
\vspace{1mm}
-\gamma_{\nu'}^{\dagger} \\
\vdots
\end{array}
\!\!\right] ,
\end{equation}
with
\begin{equation}
(\underline{u})_{{\bf r}\nu}=u_{{\bf r}\nu}\, ,
\hspace{5mm}
(\underline{v})_{{\bf r}\nu}=v_{{\bf r}\nu} \, .
\end{equation}
We now impose the condition that $\gamma_{\nu}$
as well as $\phi_{\bf r}$ obey
the Bose statistics.
This implies
\begin{equation}
\hat{\tau}_{3}=[\vec{\phi},\vec{\phi}^{\,\dagger}]
=\hat{u}\,[\vec{\gamma},\vec{\gamma}^{\,\dagger}]\,\hat{u}^{\dagger}
=\hat{u}\hat{\tau}_{3}\hat{u}^{\dagger} \, ,
\end{equation}
where $\hat{\tau}_{3}$ is an infinite matrix defined by
\begin{equation}
\hat{\tau}_{3}\equiv \left[
\begin{array}{cc}
\vspace{1mm}
\underline{1}& \underline{0}
\\
\underline{0}& -\underline{1} 
\end{array}
 \right] ,
\end{equation}
with $\underline{1}$ and $\underline{0}$ denoting the unit and the zero 
matrices, respectively.
We also require that $\hat{u}$ be composed of eigenstates of a
matrix $\hat{H}$ as
\begin{equation}
\hat{H}\hat{u}=\hat{u}\hat{E} \, ,
\label{H-u}
\end{equation}
where $\hat{E}$ is a real diagonal matrix.
Multiplying Eq.\ (\ref{H-u}) by
$\hat{u}^{-1}=\hat{\tau}_{3}\hat{u}^{\dagger}\hat{\tau}_{3}$
from the left, we obtain 
\begin{subequations}
\begin{equation}
\hat{\tau}_{3}\hat{u}^{\dagger}\hat{\tau}_{3}\hat{H}
\hat{u}\!=\! \hat{E}\, .
\label{eq1}
\end{equation}
We next take the Hermitian conjugate
of Eq.\ (\ref{eq1}) and multiply the resulting equation 
by $\hat{\tau}_{3}$ from
both sides; this yields
\begin{equation}
\hat{\tau}_{3}\hat{u}^{\dagger}\hat{H}^{\dagger}\hat{\tau}_{3}
\hat{u}\!=\! \hat{E} \, .
\label{eq2}
\end{equation}
\end{subequations}
Subtracting Eq.\ (\ref{eq2}) from Eq.\ (\ref{eq1}) and 
multiplying the resulting equation by
$\hat{\tau}_{3}\hat{u}$ and $\hat{\tau}_{3}\hat{u}^{\dagger}$
from the left and the right, respectively,
we obtain
\begin{equation}
\hat{H}^{\dagger}=\hat{\tau}_{3}\hat{H}\hat{\tau}_{3} \, .
\label{Hdagger-H}
\end{equation}
Thus, $\hat{H}$ is necessarily non-Hermitian
with the property (\ref{Hdagger-H}).

\section{Deriving Eqs.\ (\ref{Omega-eq}) and (\ref{S})}

We here derive expressions of 
the equilibrium thermodynamic potential
(\ref{Omega-eq}) and the entropy (\ref{S}).

Using Eqs.\ (\ref{hatH}) and (\ref{hatH-exp}), 
the logarithmic term in Eq.\
(\ref{LWF}) can be transformed as
\begin{eqnarray}
&&\hspace{-7mm}
L\equiv \frac{T}{2} \sum_{n} {\rm Tr}\,
\hat{1}(z_{n})\ln \bigl(\hat{\tau}_{3}K\!+\!\hat{\Sigma} 
\!-\! z_{n}\hat{1}\bigr)
\nonumber \\
&& \hspace{-3.5mm}
= \frac{1}{2}\int\!d{\bf r}
\!\int_{C} \frac{dz}{2\pi i}{\rm e}^{z0_{+}} n(z)
\sum_{\nu}\bigl\{|u_{\nu}({\bf r})|^{2}
\bigl[
\ln(E_{\nu}\!-\!z)
\nonumber \\
&&\hspace{-1mm}
+\ln(z\!-\! E_{\nu})\bigr]-|v_{\nu}({\bf r})|^{2}\bigl[
\ln(E_{\nu}\!+\!z)\!+\!\ln(-z\!-\! E_{\nu})\bigr]
\bigr\} \, ,
\nonumber \\
\label{L-def}
\end{eqnarray}
with $n(z)\!\equiv\!1/({\rm e}^{z/T}\!-\! 1)$.
Here we have used the standard procedure to 
change the summation over $n$ into a contour integral\cite{FW71}
with the replacements ${\rm e}^{z_{n}0_{+}}\!\rightarrow\! 
n(z){\rm e}^{z0_{+}}$ and
${\rm e}^{-z_{n}0_{+}}\!\rightarrow\! 
-n(-z){\rm e}^{-z0_{+}}$ in the matrix $\hat{1}(z_{n})$
of Eq.\ (\ref{1_n}).
Contour $C$ consists of two parallel lines running along the imaginary axis
as shown in Fig.\ 25.4 of Ref.\ \onlinecite{FW71}, 
which can be deformed into contour
$C'$ in the same figure along the real axis. 
Performing integration by parts in terms of $z$
and collecting residues within $C'$,
we obtain
\begin{equation}
L=\sum_{\nu}\left[T\ln(1-{\rm e}^{-E_{\nu}/T})-E_{\nu}\int\!
|v_{\nu}({\bf r})|^{2}d{\bf r}\right] \, .
\label{sum1}
\end{equation}
Next, we rewrite $\Phi$ of Eq.\ (\ref{Phi})
by using Eqs.\ (\ref{hatSigma})-(\ref{rho}) as 
\begin{eqnarray}
&&\hspace{-8mm}
\Phi=\frac{1}{4}{\rm Tr}\int\!d{\bf r}\int\!d{\bf r}'\,
\hat{\Sigma}({\bf r},{\bf r}')\biggl[
-T\sum_{n}\hat{G}({\bf r}',{\bf r};z_{n})
\hat{1}(z_{n})
\nonumber \\
&&\hspace{-1mm}
+\hat{\tau}_{3}\vec{\Psi}({\bf r}')\vec{\Psi}^{\dagger}({\bf r})
\biggr] \, .
\label{Phi-A}
\end{eqnarray}
Finally, the first integral in Eq.\ (\ref{LWF}) is transformed 
with Eq.\ (\ref{GP}) into an expression without $K$.
Substituting the result together with Eqs.\ (\ref{sum1})
and (\ref{Phi-A}) into Eq.\ (\ref{LWF}), we arrive at Eq.\ (\ref{Omega-eq}).

We next derive Eq.\ (\ref{S}) by closely following
the procedure of Ref.\ \onlinecite{Kita99}.
We first transform all the summations over $n$
in Eq.\ (\ref{LWF}) into contour integrals of using $n(z)$.
With the properties $\delta\Omega/\delta \Psi^{*}\!=\! 0$
and $\delta\Omega/\delta \hat{G}\!=\!\hat{0}$,
the differentiation $-\partial \Omega/\partial T$
need be carried out only with respect to the 
explicit $T$ dependence in $n(z)$.
The term from $\Phi$ cancels that from
the second term in the square bracket of
Eq.\ (\ref{LWF}).
The remaining term is $L$ of Eq.\ (\ref{L-def}),
whose explicit $T$ dependence is transformed 
into the first term in Eq.\ (\ref{sum1}).
We hence obtain Eq.\ (\ref{S}).



\begin{thebibliography}{99}
\bibitem{comment-exp}
There seems to be a tiny correction $\propto y\ln y$ 
in the round bracket of Eq.\ (\ref{n0-1}). However, 
it can be neglected for a wide range of 
$\delta\!\ll\! 1$ due to the smallness of the proportionality constant.
\end{thebibliography}
\end{document}